\documentclass[12pt,a4paper]{article}
\usepackage{epsfig}
\usepackage{amssymb,amsmath,amsfonts}

\begin{document}
\date{}
\title{Analysis of Multipoint Correlations in Direct Numerical Simulation}


\author{Bernhard Stoevesandt$^*$, Andrei Shishkin$^{\dag}$,\\ Robert Stresing$^*$, Claus Wagner$^{\dag}$ and Joachim Peinke$^*$\\[1mm]
{\small $^*$University of Oldenburg, Carl-von-Ossietzky-Str. 9-12, 26129 Oldenburg}\\
{\small $^{\dag}$DLR-G\"ottingen, Bunsenstr 10 37073, G\"ottingen }
}

\maketitle

\section*{Abstract}
 We examine the Markov properties of the three velocity components of a turbulent flow generated by a
DNS simulation of the flow around an airfoil section. The spectral element code Nektar has been
used to generate a well resolved flow field around an fx79w-151a airfoil profile at a Reynolds
number of Re=5000 and an angle of attack of $\alpha = 12^o$. Due to a homogeneous geometry in the spanwise direction, a Fourier expansion has been used for the third dimension of the simulation.\\
In the wake of the profile the flow field shows a von Karman street like behavior with the vortices
decaying in the wake which trigger a turbulent field. Time series of the 3D flow field were
extracted from the flow at different locations to analyze the stochastic features. In particular the existence of Markov properties in the flow have been shown for different cases in
the surrounding of the airfoil. This is of basic interest as it indicates that fine structures of turbulence can be replaced by stochastic processes. Turbulent and Markovian scales are being determined in the turbulent field and limits of standard Gaussian Langevin processes are being determined by the reconstruction of a flow field in time and space.

\section{Introduction}
Although computational methods for fluid dynamics have made enormous progress during the recent years, the problem remains to resolve fine structures of highly turbulent flow. New turbulence models have improved the turbulence simulations to quite some extend \cite{richez08}\cite{kim06}. Nevertheless, even with the application of new techniques, it remains difficult to correctly simulate turbulent loads e.g. for wind turbines \cite{bechmann09}\cite{sezer-uzol06}. Already the accurate simulation of the flow around single airfoils at higher Reynolds numbers is a difficult task \cite{breuer07}\cite{LESFOIL} as it is difficult to grasp the correct turbulence properties everywhere in the flow. Therefore a lot of effort has been made to improve the modeling of the turbulence in small scales with subgrid models. The most common models are so called ``deterministic'' models based on eddy viscosities. Meneveau et. al. give a good overview on such models for large eddy simulations (LES) \cite{meneveau00}.\\
Since a deterministic description of turbulence still has its limits, it is very convenient to describe the turbulence a time series \cite{muschinski04}. The reconstruction of stochastic time series has made quite some progress in the recent years. First approaches involved the Fokker-Planck equation with Gaussian diffusion \cite{friedrich97}. However today the research expands to Langevin equations for Levy processes \cite{lubashevsky09} or even non-Markovian fields \cite{farias09}. From this evolved  approaches to model turbulent data in different fields of science. Hence such approaches arouse some interest for the modeling of turbulent flows in numerical flow simulations \cite{Monin75}.\\
Some approaches to use stochastic models for flow simulations have been undertaken. Laval and Dubrulle proposed a Langevin approach for LES models which showed promising first results. Based on the Rapid Distortion Theory (RDT)-model, described in \cite{laval2006}\cite{dubrulle04} and \cite{laval2003}, they developed a Langevin equation for an LES model, with friction made of viscosity and rapid distortion by resolved scales, using stochastic forcing with a mean value generated from the energy cascade. The model was validated on the flow in an empty cube against a DNS simulation.\\
Mostly however, stochastic models were developed and used in the context of particel tracking or a Lagrangian framework as described by Bakosi, Pope, Shotorban or Fox  \cite{bakosi08}\cite{pope98}\cite{shotorban06}\cite{fox95} \cite{fox03}. The particle tracking method by Bakosi works with a finite element grid in a Eulerian framework in contrast to the work of other groups. The physical magnitudes of the particles are described by probability density functions (pdfs) at each specific point.\\
This approach requires the knowledge of the pdfs within a flow. One way to obtain these pdfs is the reconstruction by the Fokker-Planck equation using time series data. In an isotropic and homogeneous field of Gaussian turbulence this method is very easily applied and valid. In non-homogeneous flow fields the stochastic properties of the flow differ differ. A stochastic reconstruction of the fields using the method proposed by Friedrich and Peinke in \cite{friedrich97}\cite{friedrich97b} might nevertheless still be a promising approach to gain knowledge about the turbulent characteristics of a flow field for the modeling of pdfs. This shall be the focus of the following contribution.\\

\section{The underlying equations for turbulence models}
The starting point for stochastic modeling is to consider a quantity in a statistical context. In fluid dynamics this is done using the Reynolds decomposition 
\begin{equation}
   u = \overline{u} + u'
   \label{Rey-Decomp}
\end{equation}
where $\overline{u}$ is the averaged velocity field and $u'$ denotes it's fluctuation. For time averaging the decomposition applied to the Navier-Stokes equations leads to the so called Reynolds Averaged Navier Stokes equations (RANS). For incompressible fluids they read:
\begin{equation}
	\partial_t \overline{u}_i + \partial_j \overline{u}_j \overline{u}_i + \partial_j \overline{u_j' u_i'} = 
	-\frac{1}{\rho} \partial_i \overline{p} + \nu \partial^2_j \overline{u}_i.	
	\label{RANS}
\end{equation}
The incompressible continuity equation is then given as
\begin{equation}
   \nabla \cdot (\overline{u} + u') = \nabla \cdot \overline{u}=\nabla \cdot u'= 0.
\end{equation}
To solve the closure problem, models for the Reynolds stress term 
\begin{eqnarray}
\partial_j \overline{u_j' u_i'}
\label{RStress}
\end{eqnarray}
are needed.\\
This is getting more complicated for LES models. Here the velocity field is spatially filtered and split into a resolved field $\overline u_i$ and a subgrid field given by $u'_i=\overline{u}_i-u_i$ for $u_i(x_i,t)$ (note that here the average $\overline{u}_i$ describes an average over a spatial region and not - like in RANS - in time or over ensembles). In this situation the momentum equation of the Navier-Stokes equations turn out to be
\begin{eqnarray}
       \partial_t \overline{u_i} + \overline{(\overline{u}_i \partial_i )\overline{u}_j} + \overline{(\overline{u}_i \partial_i)u'_j} + \overline{(u'_i \partial_i)\overline{u}_j}+\partial_j \overline{u_j' u_i'} = -\frac{1}{\rho} \partial_i \overline{p} + \nu \partial^2_j \overline{u}_i .
      \label{LES-eq}
\end{eqnarray}
In this case a model for the forcing term $l=(\overline{u}_i \partial_i)u'_j + (u'_i \partial_i)\overline{u}_j$ is needed in addition to the subgrid scale stress tensor. Recently, Laval et al. \cite{laval2006} propose a stochastic approach to the first unknown term by putting up a Langevin equation:
\begin{eqnarray}
       \partial_t l = Al + \xi,
\label{Laval_Langevin}
\end{eqnarray}
where a Gaussian noise $\xi$ needs to be added in the equation, while $A$ is a generalized evolution operator. Based on the observation that subfilter-scales are mostly dependent to resolved scales by a linear process similar to rapid distortion, Laval et al. derive from the momentum equation a model for the evolution operator $A$ \cite{laval2006}.\\
Another approach is the one by Bakosi et.al. \cite{bakosi08} who model the stress terms using a (Gaussian) Wiener process to gain a pdf for the stresses. However the knowledge of the pdfs and the ability of their reconstruction are a premise for such a method. This requires the correct description of the turbulent field at all scales. Here the n-point correlation approach by using increments and the knowledge about Markov properties in a flow field is a basis \cite{renner01}\cite{boettcher06} from which further modeling can proceed.
 Tutkun et. al. have done an analysis of a spatial flow field experimentally \cite{tutkun04}.\\
As a typical example of a problem of turbulent flow we analyze here a flow field obtained by a DNS simulation of the flow around an airfoil to describe the method. The simulation, its parameters and the resulting field are presented at first in the following section. Then the basic aspects of our stochastic approach is explained. Finally the results of the analysis of the data are discussed.

\section{The Simulation} \label{simSetup}
\subsection{Simulation Parameters}
\begin{figure}[htb]
  \includegraphics[width=.45\linewidth]{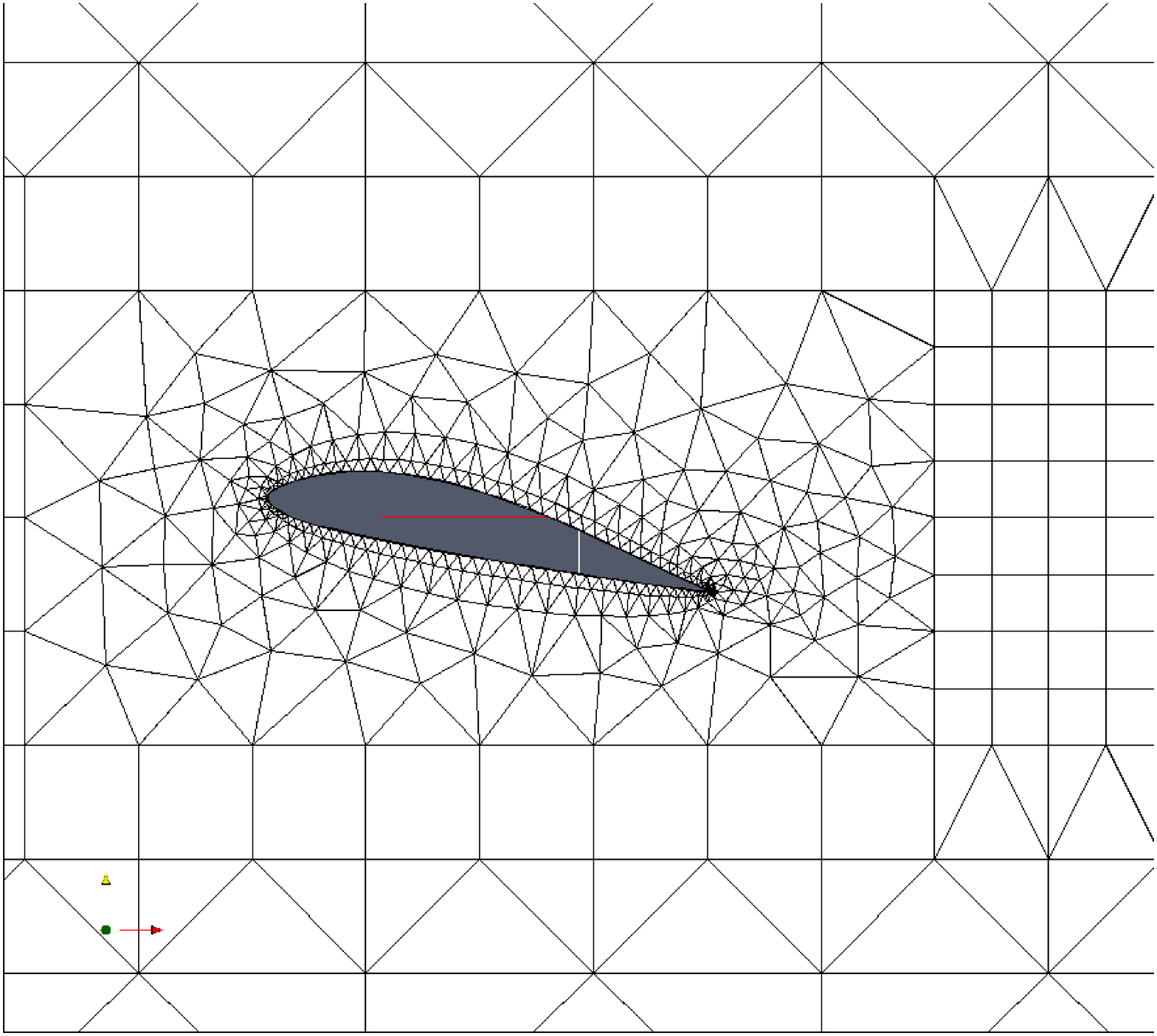}
  \includegraphics[width=.45\linewidth]{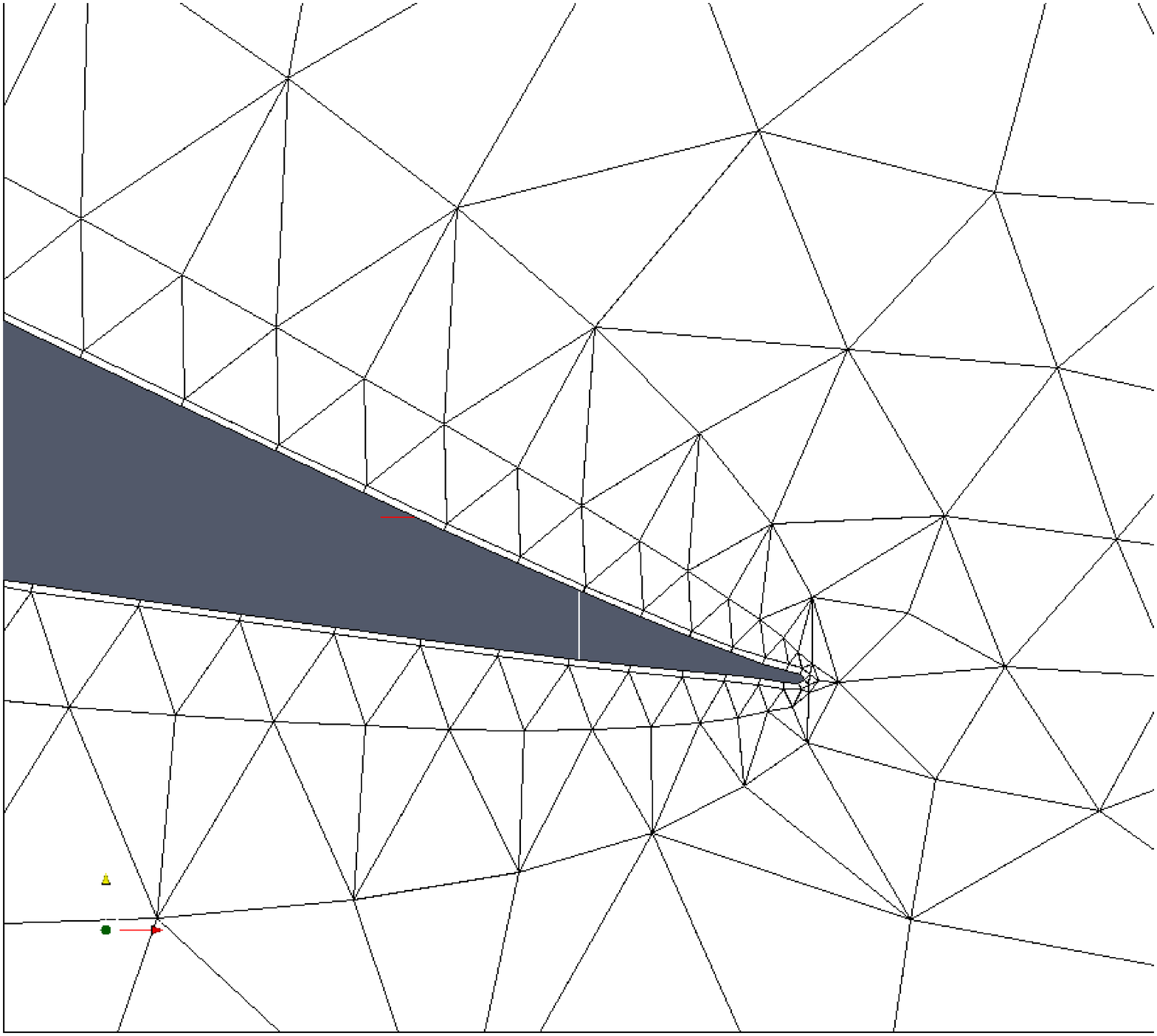}
  \caption{2D grid of the airfoil setup, consisting of 1179 quadrilaterals and 937 triangles. A close-up on the right shows the boundary resolution and the resolution of the tail.}
  \label{fig.mesh}
\end{figure}

 To gain a time series of a turbulent flow field a DNS simulation has been done using the high order spectral element code $\cal N \varepsilon \kappa \cal T \alpha${\it r} \cite{kirby99}. It combines finite element methods with the accuracy of spectral methods using Jacobi polynomials for the spectral expansion \cite{sherwin_karniadakis2005}. For the velocity pressure coupling a stiffly stable pressure correction scheme and an Adams-Bashfort predictor corrector time step were used as described in \cite{Karniadakis_Israeli}.\\
To obtain a well-studied and representative turbulent flow field from numerical flow simulations, the flow over an airfoil has been investigated \cite{stoevesandt06}. The configuration is a 3D flow over a section of an fx79w-151a airfoil at an angle of attack of $\alpha=12^\circ$ at a Reynolds number of $Re=5000$ in respect to the chord length.\\
 The used $\cal N \varepsilon \kappa \cal T \alpha${\it r} code works with a 2D mesh with rather few elements for a first spatial discretisation. Further, the interior of the elements is calculated using an expansion by Jacobi polynomials \cite{warburton}. Here we used a polynomial order of n=9. The discretisation in the homogeneous spanwise direction of the airfoil was realized using the ``Fourier'' version of the code which calculates the third - spanwise - dimension using a Fourier expansion in combination with periodic boundary conditions on the sides of the domain (for validations see e.g. \cite{Ma}\cite{Dong}).\\
Measured in chord lengths $L_c$, the domain size expands from $-10L_c$ to $10L_c$ in the cross-flow direction and from $-6L_c$ to $20L_c$ streamwise direction. The domain size in spanwise direction was set to $\pi \cdot L_c$, which showed good results in a simulation on the flow around a cylinder at Re=3900 \cite{Ma}. The extension in the spanwise direction seems to be sufficient, as integral scales in the turbulent region in all directions are one order smaller than the domain size (see $L$ in table \ref{spacing}). 64 Fourier planes have been used for the expansion in spanwise direction together with a 2D hybrid mesh consisting of 2116 elements (see fig.\ref{fig.mesh}).\\
To resolve the boundary layer flow a quadrilateral grid layer was created around the airfoil. The original shape of the airfoil was not changed since every detail of the geometry has been taken into account. As the tail of the airfoil had a tiny triangular shape, the resolution of this tail leads to very small elements at the trailing edge (see fig.\ref{fig.mesh} on the right). This resulted in very small time steps of order $t \approx 10^{-5}$. In our simulation the maximum CFL was $\le 0.4$.\\
To avoid instabilities which in turbulent flows often develop from Neumann outflow boundary conditions a sponge layer has been implemented increasing the viscosity in the last $3 L_c$ before the outflow. Finally we used a uniform laminar velocity distribution at the inflow.

\subsection{Simulation Results}
\begin{figure}[htb]
a) \includegraphics[width=.47\linewidth]{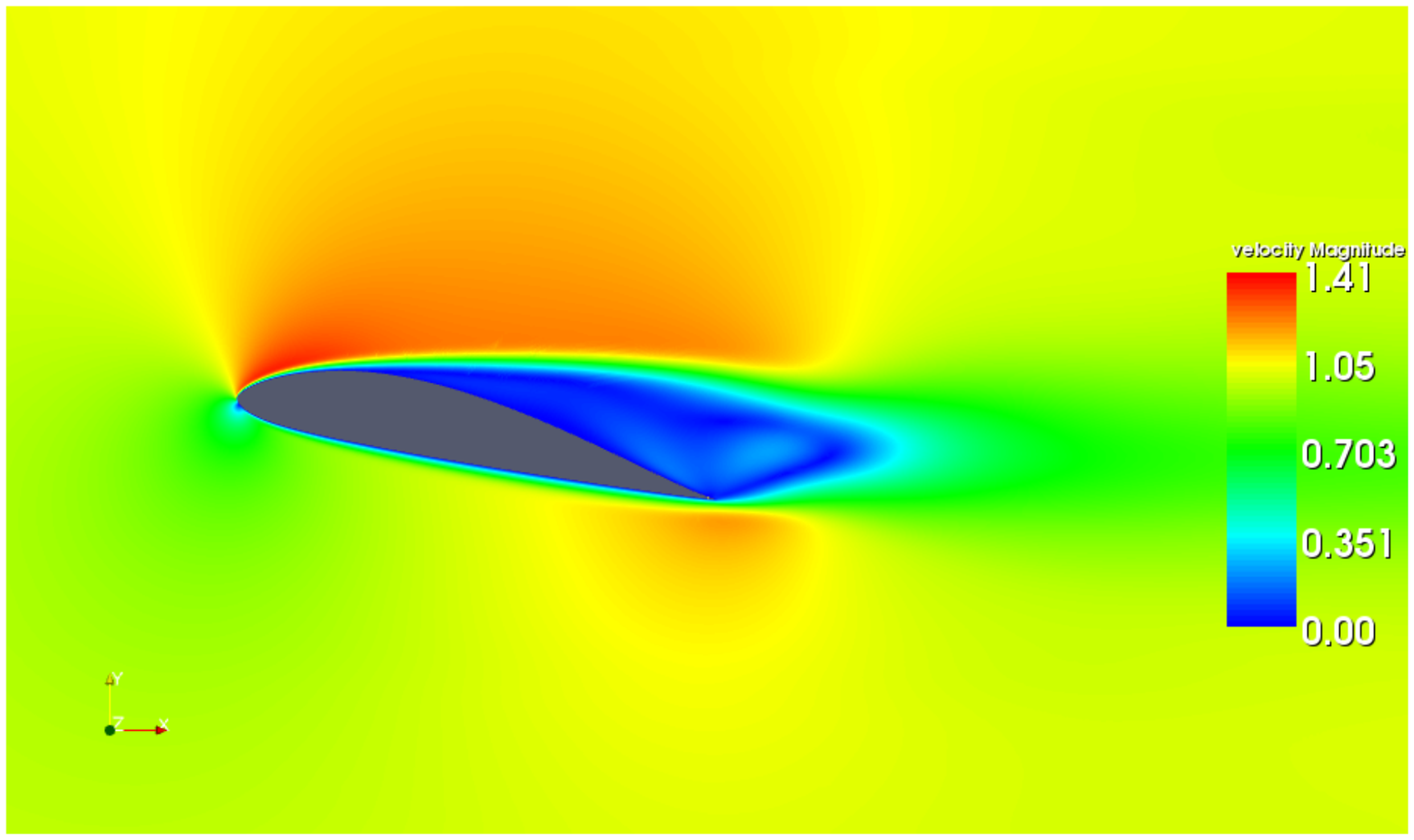}
b) \includegraphics[width=.47\linewidth]{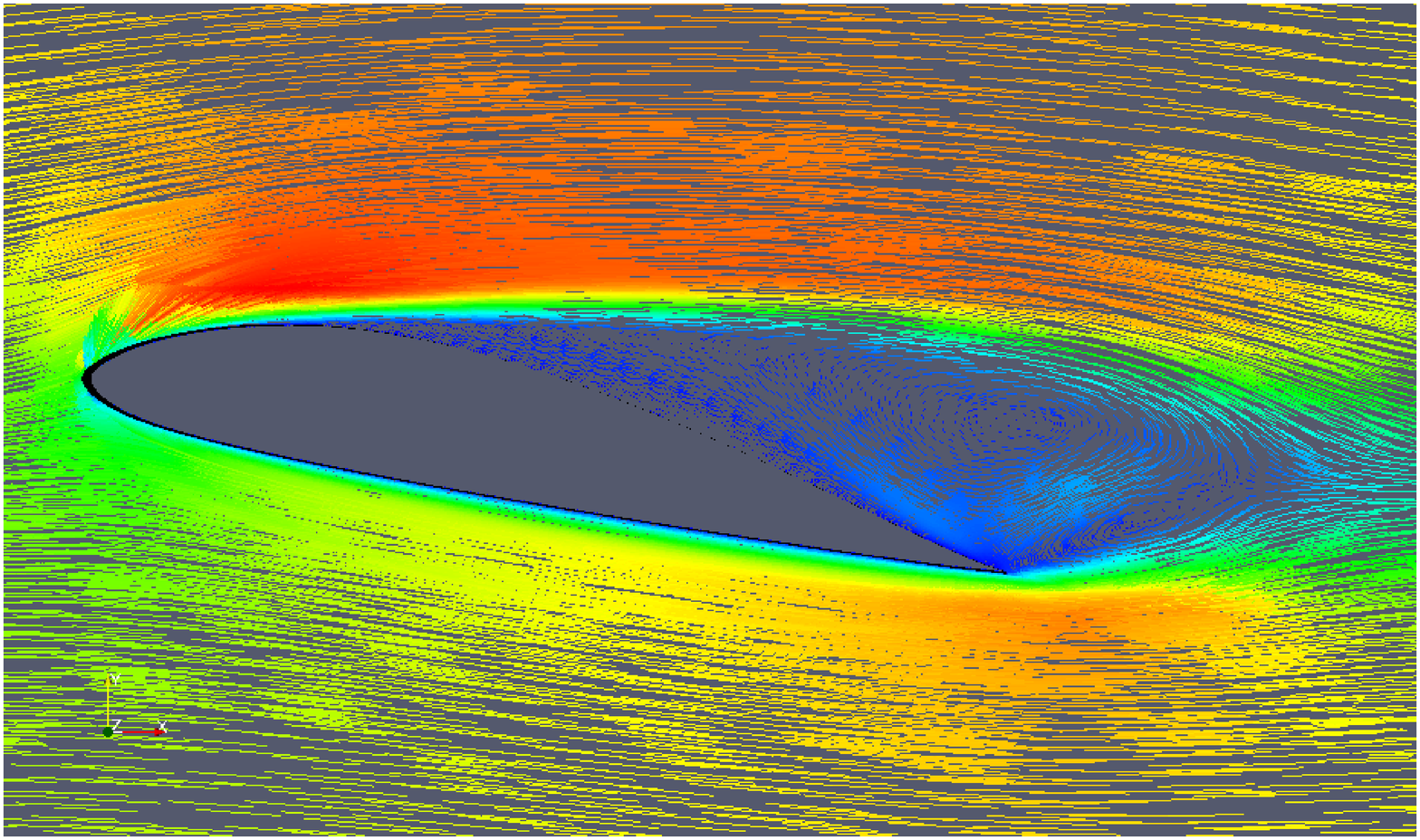}
  \caption{Mean velocity of the flow field over t=20 in normalized time scales. The magnitudes of the velocity given in a) by the colors, additionally the vector orientation of the flow in b).}
  \label{av-flow}
\end{figure}

\begin{figure}
a) \includegraphics[width=.47\linewidth]{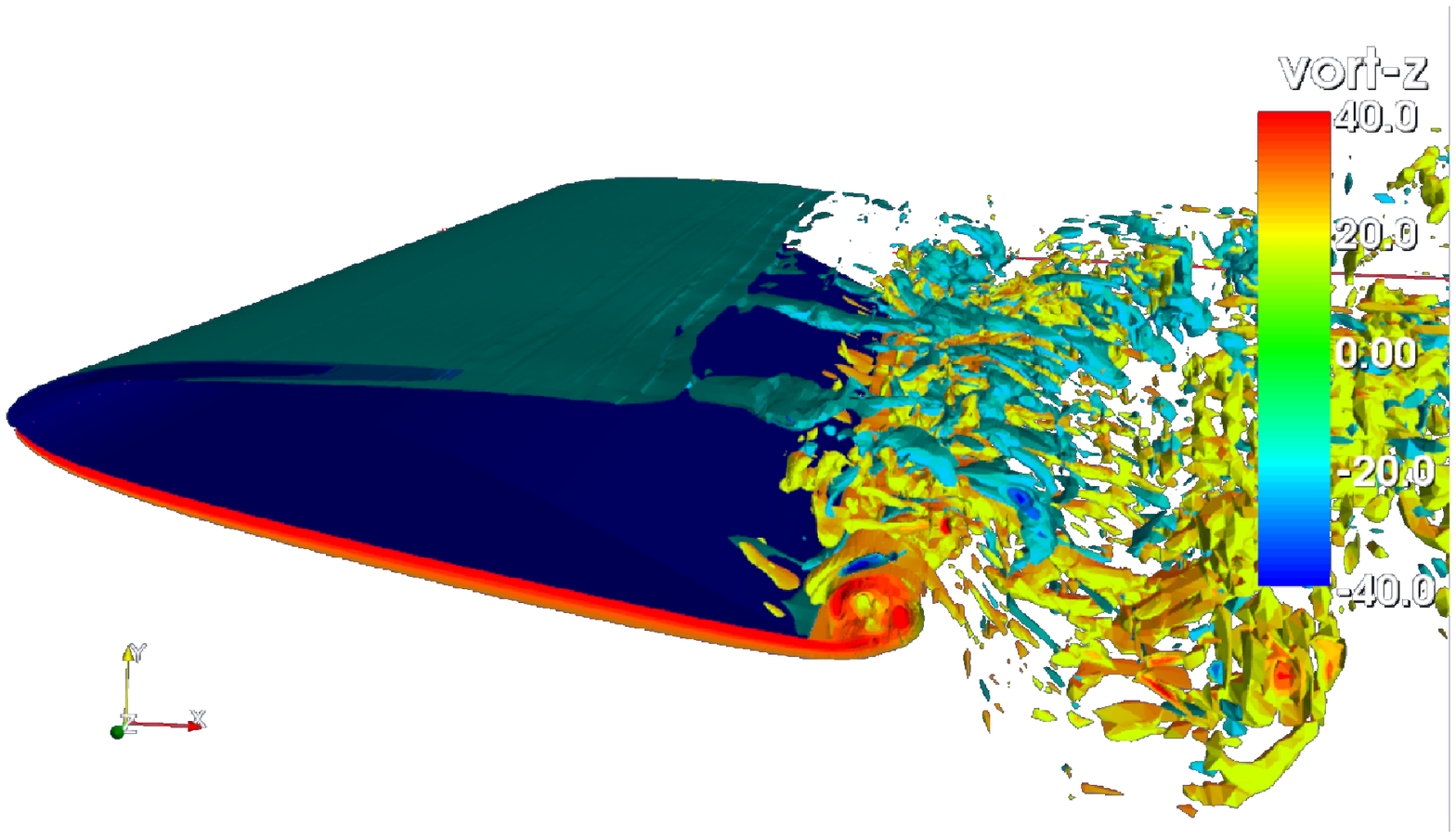}
b) \includegraphics[width=.47\linewidth]{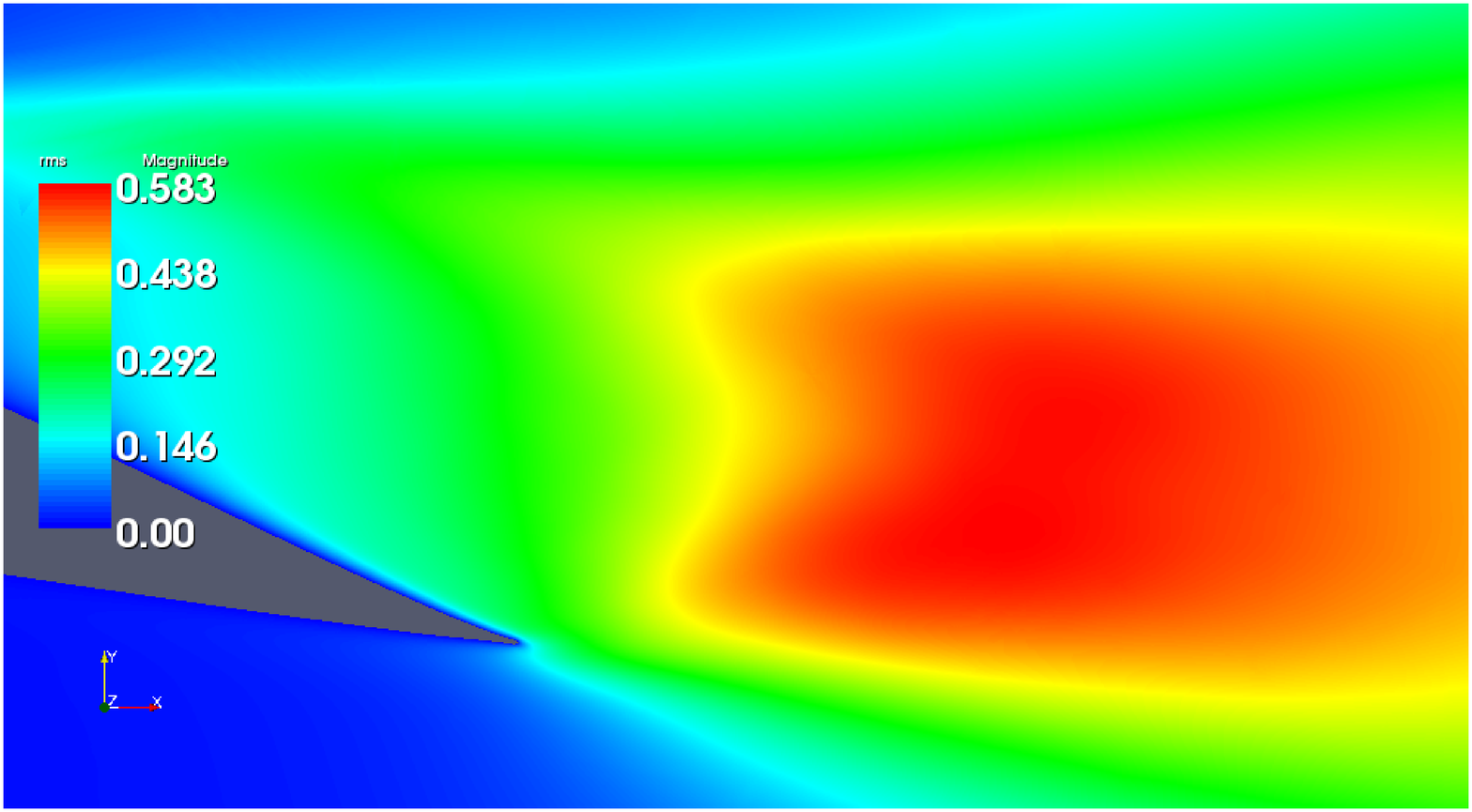}
  \caption{Snapshot of vorticity components in z-direction a) and rms-fluctuations of the velocity magnitude in b).}
  \label{fig.res}
\end{figure}

\begin{figure}
  \includegraphics[width=.60\linewidth]{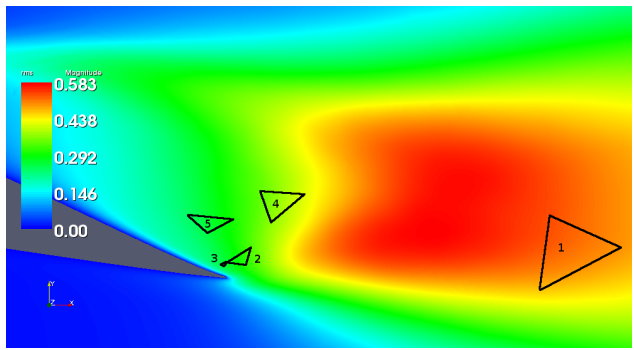}
  \caption{Selected grid in the region just above the trailing edge for the further investigation of the turbulence}
  \label{rms-elem}
\end{figure}

Fig. \ref{av-flow} depicts the average flow field over the airfoil. Here and further on, all units are dimensionless and normed to chord length of the airfoil. The average flow field is mainly characterized by a laminar flow separation in the tip region with an impingement point in the area behind $\frac {4}{5} L_c$. The shear flow which develops downstream the separation point leads to a von Karman like vortex street in the rear part of the airfoil. Triggered by the vortices a turbulent wake flow field appeared (see e.g. vorticity contours in fig. \ref{fig.res}a)). To give an impression of the distribution of the turbulence intensity in the field, fig. \ref{fig.res}b) shows contours of the RMS of the velocity magnitude of the flow in the vicinity of the airfoil. In this figure high fluctuations in the velocity magnitude  are recognizable mainly close to the trailing edge and in the wake. Thus for the study of turbulence at and around the airfoil the regions at the trailing edge and in the wake of the airfoil were of main interest.\\
To perform a time series analysis of the flow properties specific points effected by the turbulent wake flow have been chosen within selected elements of the mesh in the region depicted in fig. \ref{rms-elem}.\\
The regions selected were positioned on one hand in the wake (element 1), on the other hand close to the trailing edge of the airfoil at different heights (elements 2-5). It must be noted that all position selected for the recording of time series are located in regions with a triangular unstructured grid. Within the selected elements at the spanwise position of $z=2.11 L_c$ a time series of 10922 datasets at 208 different spatial points were collected at a sampling rate of $f_r=400$. Out of the five selected elements 62 points have been chosen for the purpose of the following analysis.\\

\section{The Definition of Markov Properties} \label{Markov_Def}
In the following the stochastic analysis of the turbulent flow field will be discussed. The final aim is to gain more knowledge of the Reynolds stress tensor, which for a the stochastic analysis reads
	\begin{equation}
	  \overline{u'_{i} u'_{j}}=\int u'_{i} u'_{j} p(u'_{i} u'_{j})d(u'_{i} u'_{j}),
	  \label{Stress}
	\end{equation}
where $p(u'_iu'_j)$ is the probability density function (pdf) of the components of the Reynolds stress tensor $u'_i u'_j$.\\ 
In the following we will generalize the discussion on the stochastic features of important fluctuating quantities such as Reynolds stress. For turbulence it is of central interest to characterize the spatial correlations of quantities $\phi(x_i,t)$. Therefore it is common to investigate the statistics of increments in a scale of $r$ at the same point in time $t$
\begin{eqnarray}
  \phi(r,t)=\phi(x_i-r,t)-\phi(x_i,t),
  \label{increment}
\end{eqnarray}
where $x_i$ denotes the position in space and $t$ is the time.\\
The following analysis will be based on probabilities of such increments at different scales. The conditioned probability function of $\phi$ can be described as
\begin{equation}
p(\phi_1,r_1|\phi_2,r_2;...;\phi_n,r_n)=\frac{p(\phi_1,r_1;\phi_2,r_2;...;\phi_n,r_n)}{p(\phi_2,r_2;...;\phi_n,r_n)}
\label{cpdf}
\end{equation}
where the left hand side reflects the probability density function of the magnitude $\phi_1$ at the incremental scale $r_1$ under the condition that at the scales $r_2,...,r_n$ the magnitudes $\phi_2,...\phi_n$ have selected fixed values. Here we use $r_i < r_{i+1}$ and the abbreviation $\phi_1=\phi(r_1,t)$.\\
	In stochastics we speak of Markov properties, if the pdf of a dataset can be completely described with the knowledge of the neighboring magnitude at scale $r_{n-1}$ without being influenced by the next magnitude at $r_{n-2}$. In case of the increments this means:
	\begin{equation}
	    p(\phi_n,r_n|\phi_{n-1},r_{n-1};...;\phi_{1},r_{1})
	    =  p(\phi_{n},r{_n}|\phi_{n-1},r_{n-1})
	    \label{markov}
	\end{equation}
	If this is the case the general n-scale statistics can be closed by the two scale statistics $p(\phi_{n},r{_n}|\phi_{n-1},r_{n-1})$.  Furthermore, it is possible to apply the Kramers-Moyal expansion of the data series, which leads to:
\begin{eqnarray}
-r \frac{\partial p(\phi,r | \phi_0,r_0)}{\partial r} & = & \sum \limits_{k=1}^{\infty} \left ( - \frac{\partial}{\partial \phi} \right)^k D^{(k)}(\phi,r) p(\phi,r | \phi_0,r_0).
\label{Kramer-Moyal}
\end{eqnarray}
The coefficients of this expansion are defined as
\begin{eqnarray}
D^{(k)}(\phi,r)=\lim_{\Delta r \to 0} M^{(k)}(\phi,r,\Delta r)
\label{KM-coeff}
\end{eqnarray}
\begin{eqnarray}
M^{(k)}= \frac{r}{k!\Delta r}\int_{-\infty}^{\infty}(\phi'-\phi)^k p(\phi',r-\Delta r|\phi,r)d\phi'.
\label{M-coeff}
\end{eqnarray}
This definition of $D^(n)$ can be used to estimate its value directly from the data \cite{friedrich97}\cite{friedrich97b}\cite{renner01}.\\
The theorem of Pawula states that $D^{(k)}=0$ $\forall$ $k>2$ if the coefficient of the expansion $D^{(4)}= 0$ \cite{risken96}. In such a case the Kramers Moyal expansion (\ref{Kramer-Moyal}) truncates to the second order, yielding the Fokker-Planck equation (also called Kolmogorov forward equation) \cite{risken96}:
	\begin{equation}
	  \partial_r p(\phi,r|\phi_0,r_0) 
	  =  [-\partial_{\phi} D^{(1)}p(\phi,r) + \partial^2_{\phi} D^{(2)}p(\phi,r)]p(\phi,r|\phi_0,r_0).
	  \label{FP}
	\end{equation}
The first coefficient $D^{(1)}$ is referred to as the drift coefficient, the second coefficient $D^{(2)}$ is called diffusion coefficient and determines the strength of a Gaussian distributed white noise. This noise is $\delta$-correlated due to the Markov properties \cite{risken96}. The Fokker-Planck equation can be reformulated as a Langevin equation following It\^o or Stratonovich (see \cite{risken96}). The It\^o formulation is:
\begin{eqnarray}
-\frac{\partial \phi(r)}{\partial r}=\frac{1}{r} D^{(1)}(\phi,r)+\sqrt{\frac{1}{r}D^{(2)}(\phi,r)}\Gamma(r).
\label{Langevin}
\end{eqnarray}
$\Gamma(r)$ is the Gaussian distributed white noise.\\
Under these conditions it is possible to analyze the Reynolds stress tensor $\phi = u'_i u'_j$. Having determined  $D^{(1)}$ and $D^{(2)}$, the Fokker-Planck eq. (\ref{FP}) allows to reconstruct a series on all n-scales of $r_n$ \cite{nawroth06}. Thus it is possible to reconstruct $p(\phi_i,r_i)$ for given scales $r_i$, where $r_i$ may be given by a finite spatial relation. This can be used as an expression for the components of the Reynolds stress tensor $u'_iu'_j(r)$ eq. (\ref{RStress}). A similar procedure could be constructed for other relevant turbulent quantities like the additional stress terms in equation (\ref{LES-eq}).\\
The aim is now to find out under which conditions this stochastic simplification holds for a given flow problem. However, the geometrical distribution of the data points of the flow field makes a straight forward analysis difficult since not enough straight $r_n$ scales could be evaluated for a complete reconstruction. Therefore the analysis is being done using the Taylor hypothesis of frozen turbulence, which is applied quite frequently also in experimental works (like e.g. \cite{renner01}\cite{boettcher06}), by
\begin{eqnarray}
r=\tau \overline{u}.
\label{Taylor}
\end{eqnarray}
Now, $\tau=(t-t_0)$ is a time scale by which the same analysis can be undertaken as in given above in equations (\ref{cpdf})-(\ref{M-coeff}).
In this paper we focus on the most important underlying Markov property. Thus we look if
\begin{eqnarray}
p(u_0,\tau_0|u_1,\tau_1;u_2,\tau_2))=p(u_0,\tau_0|u_1,\tau_1)
\label{markov2}
\end{eqnarray}
in time and in spatial scales if
\begin{eqnarray}
p(u_0,r_0|u_1,r_1;u_2,r_2)=p(u_0,r_0|u_1,r_1)
\label{markov3}
\end{eqnarray}
holds, $u$ is one velocity component and $r=x_0-x_{1}$ is a spatial scale for this coordinate. For simplicity reasons the nomenklatura used for the coordinates and orientations has been chosen as in usual Cartesian coordinates giving $(x,y,z)$ for the streamwise, crossflow and spanwise direction as well as $(u(\tau),v(\tau),w(\tau))$ for the velocity increments in this direction.\\


\section{Analysis of the results}\label{stochAnalysis}
\begin{center}
\begin{figure}[htb]
  \includegraphics[width=.60\linewidth]{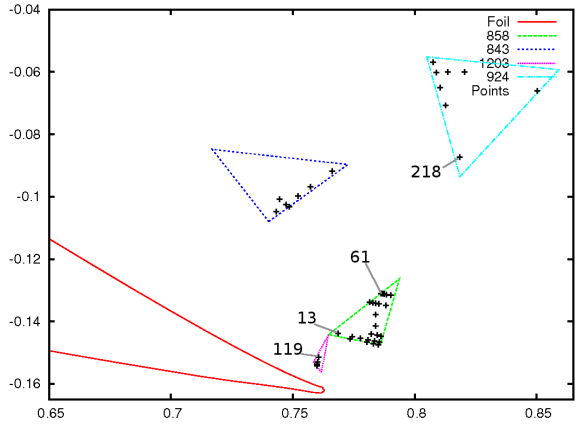}
  \caption{Geometric positioning of the evaluated points in the region over the tail of the airfoil. The numbered points have been selected for further analysis and reconstruction}
  \label{points}
\end{figure}
 \end{center}

From the over all flow field four points (marked as 13,61,119 and 218) have been selected for further evaluation (see fig.\ref{points}). In table \ref{spacing} the scale of the spacing between the points \textbf{$r$} is given for all components (x,y,z) as well as the mean velocity components $\overline{u_i}$, standard deviation $\sigma_i$ of the velocities and estimations of the integral, Taylor and dissipation length scales, represented by $L_i$, $\lambda_i$ and $\eta_i$ respectively.  All length scales have been calculated according to \cite{pope00}. Since the Taylor length is a fit with possible deviations and the dissipation scale is derived from the Taylor length, both more or less give values for the order of magnitude and are not to be seen as exact values.\\
\begin{table}[htb]
\begin{tabular}{|c|c|c|c|c|c|c|c|c|}
\hline
Point&Component&$r$&$\overline{u}$&$\sigma$&$L$&$\lambda$&$\eta$\\
\hline
218&x&0.0&-0.14&0.253&0.1805&0.0139&0.0005\\
\hline
218&y&0.0&-0.12&0.216&0.1519&0.0087&0.0004\\
\hline
218&z&0.0&-0.006&0.157&0.3308&0.0048&0.0003\\
\hline
61&x&0.032&-0.039&0.2077&0.1523&0.0032&0.0002\\
\hline
61&y&0.044&-0.074&0.171&0.1539&0.0065&0.0004\\
\hline
61&z&0.0&-0.004&0.129&0.1666&0.0033&0.0003\\
\hline
13&x&0.050&-0.012&0.183&0.1598&0.0014&0.0002\\
\hline
13&y&0.057&-0.058&0.112&0.1771&0.0064&0.0005\\
\hline
13&z&0.0&0.004&0.130&0.2850&0.0041&0.0004\\
\hline
119&x&0.058&0.028&0.151&0.1689&0.0033&0.0003\\
\hline
119&y&0.066&-0.036&0.059&0.1824&0.0033&0.0005\\
\hline
119&z&0.0&-0.001&0.135&0.5075&0.0024&0.0003\\
\hline
\end{tabular}
\caption{The analyzed points, their geometrical distances - with point 218 as a reference point -, their velocity and the statistical properties of the velocity for all components.}
\label{spacing}
\end{table}

\subsection{Analysis in time scales}\label{Results}
\begin{center}
\begin{figure}[htbp]
$\begin{array}{c@{\hspace{0.3in}}c}  
 \multicolumn{1}{l}{\mbox{\bf }}  &
 \multicolumn{1}{l}{\mbox{\bf }} \\ [-0.5cm]
\epsfxsize=2.78in
\epsffile{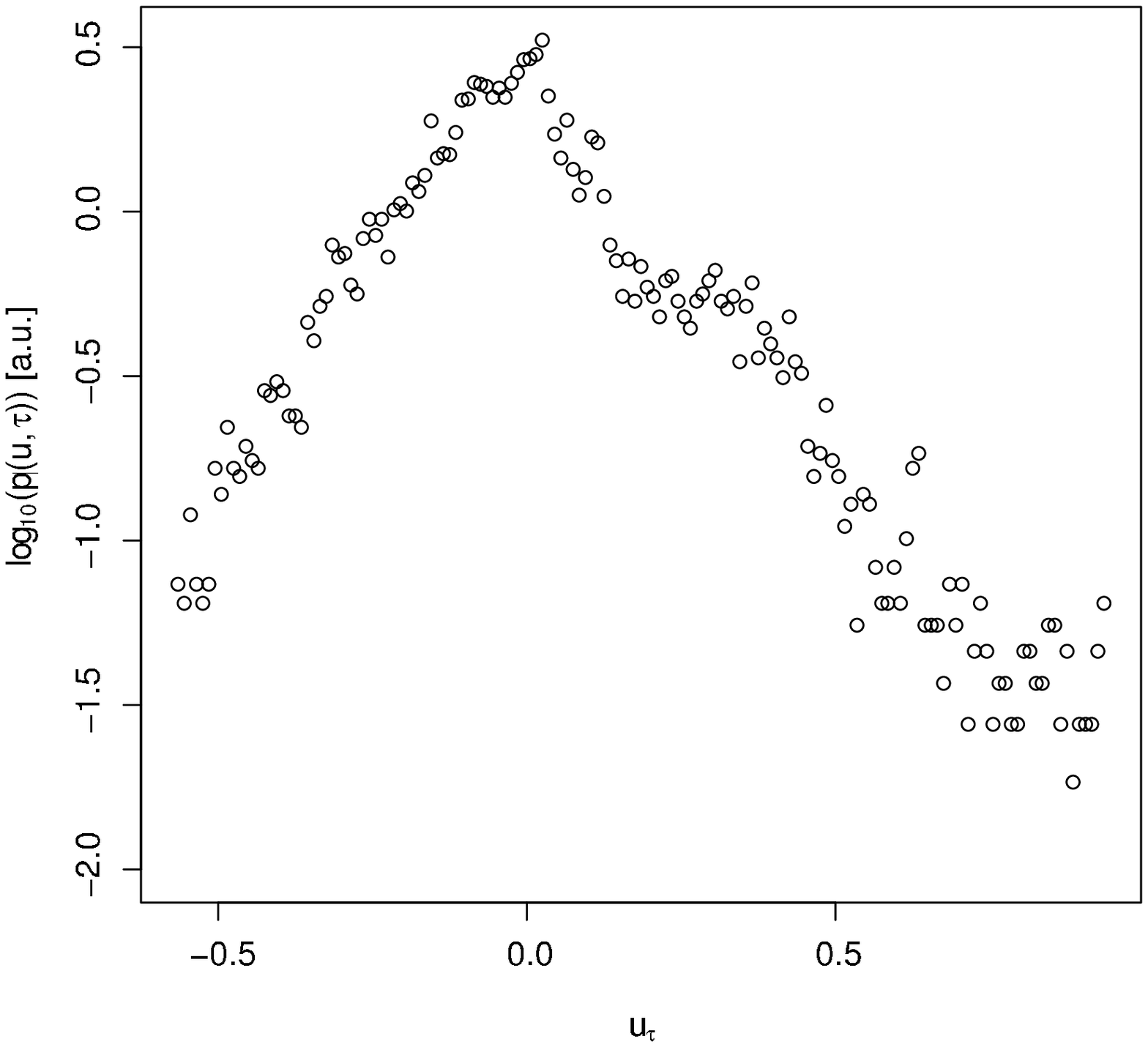}
 \put(-210,150){a)}  & 
\epsfxsize=2.78in
\epsffile{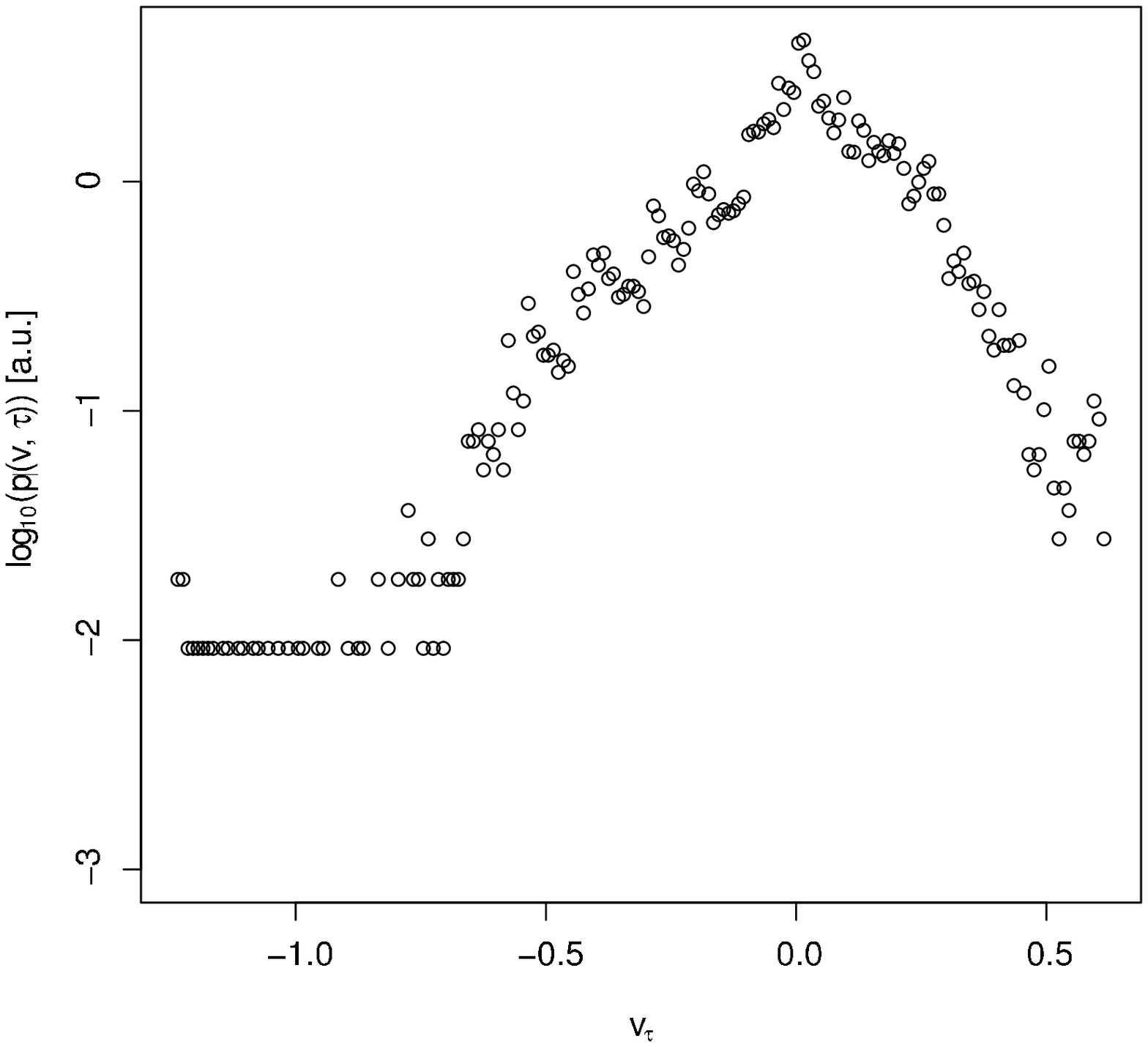}
\put(-200,150){b)}\\ [1.0cm]
\multicolumn{1}{l}{\mbox{\bf }}  &
\multicolumn{1}{l}{\mbox{\bf }}   \\ [-1cm]
\epsfxsize=2.78in
\epsffile{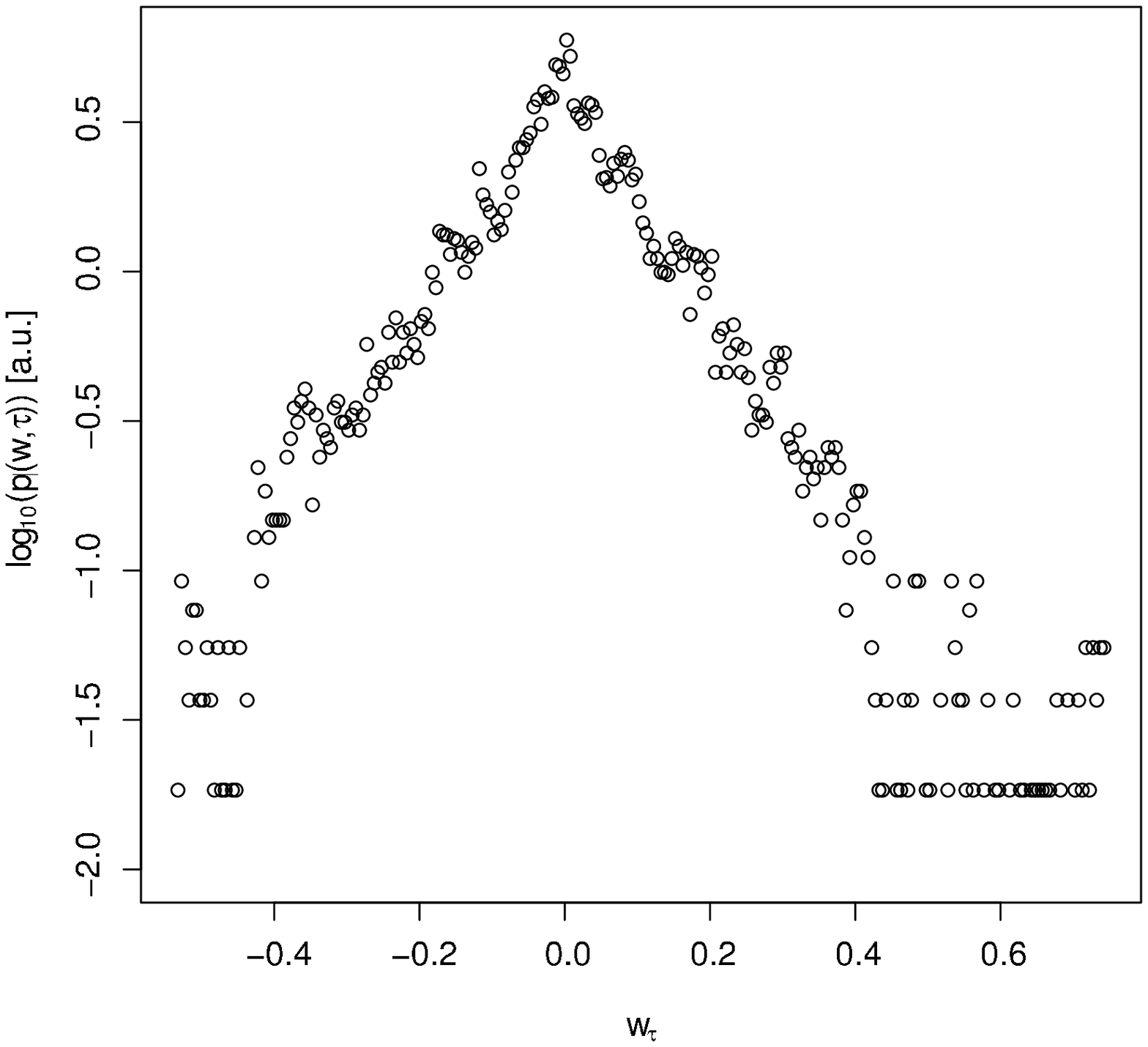}
 \put(-210,150){c)} &
\epsfxsize=2.78in
\end{array}$
\caption{Histogram of the velocity increments in semi-logarithmic presentation for u in a), v in b) and w in c) and $\tau_0=0.15$.}
\label{pdf-comp}
\end{figure}
\end{center}

\begin{center}
\begin{figure}[htbp]
$\begin{array}{c@{\hspace{0.3in}}c}  
 \multicolumn{1}{l}{\mbox{\bf }}  &
 \multicolumn{1}{l}{\mbox{\bf }} \\ [-0.5cm]
\epsfxsize=2.78in
\epsffile{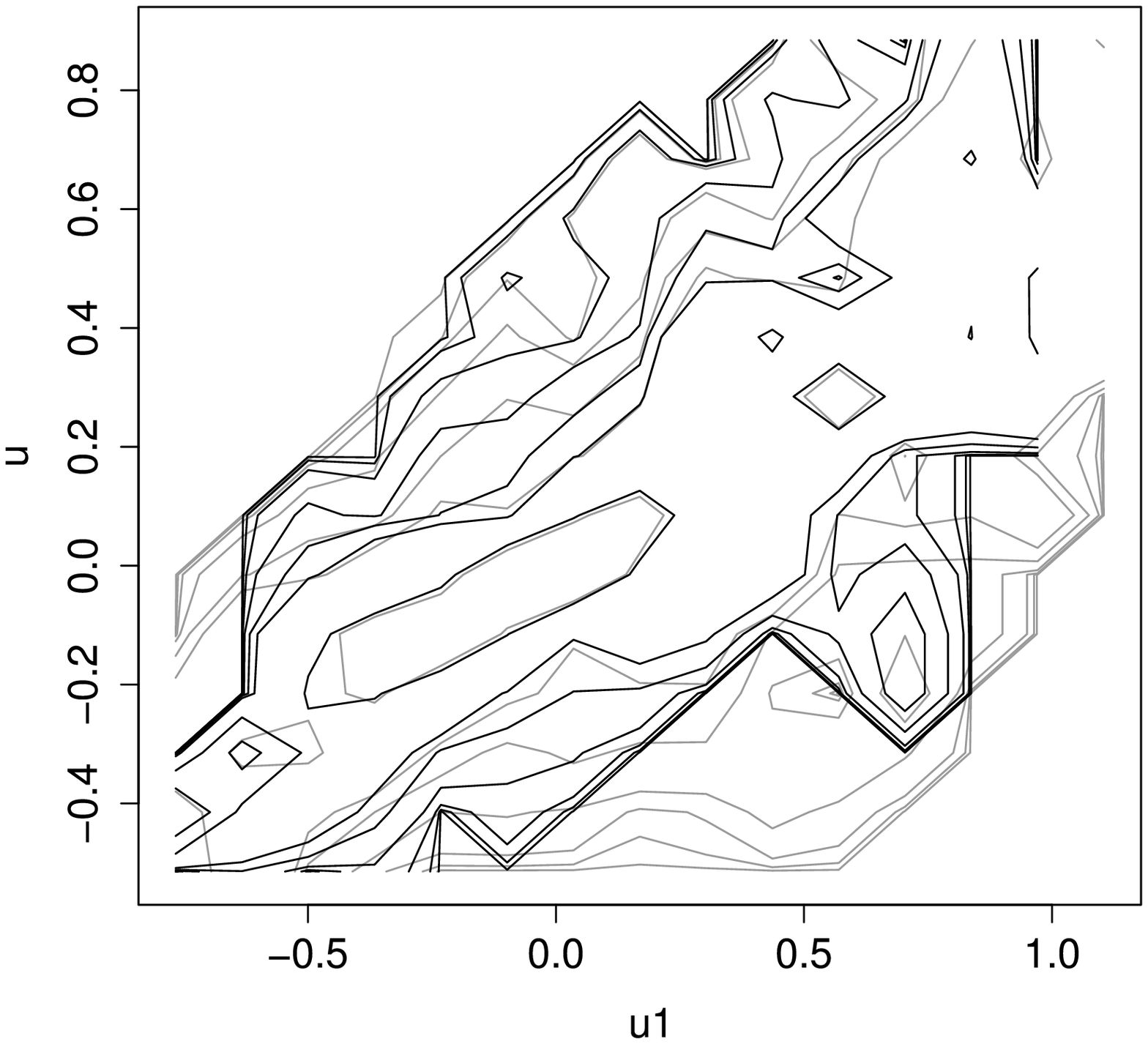}
 \put(-210,150){a)}  & 
\epsfxsize=2.78in
\epsffile{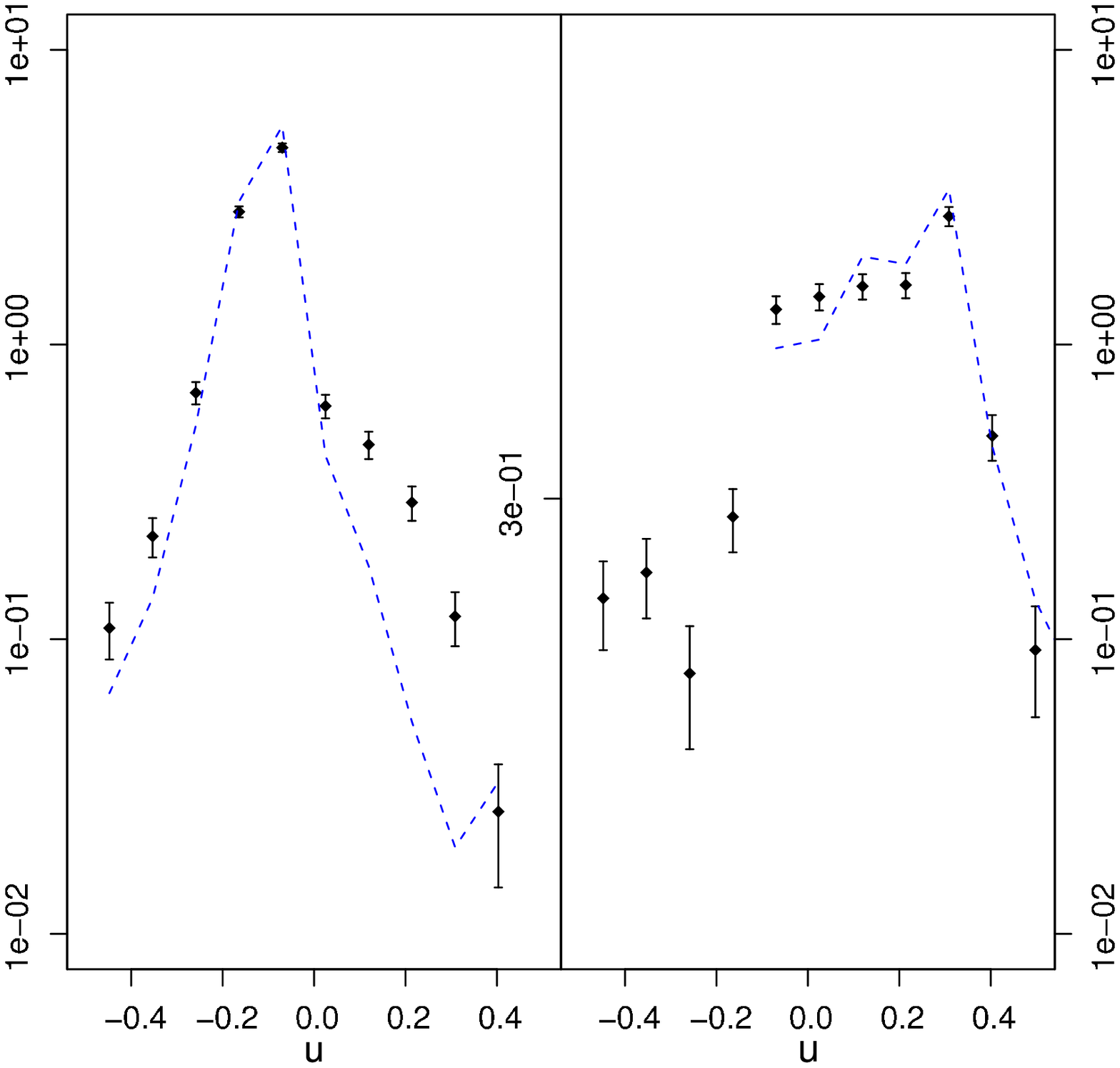}
\put(-200,150){b)}\\ [1.0cm]
\multicolumn{1}{l}{\mbox{\bf }}  &
\multicolumn{1}{l}{\mbox{\bf }}   \\ [-1cm]
\epsfxsize=2.78in
\epsffile{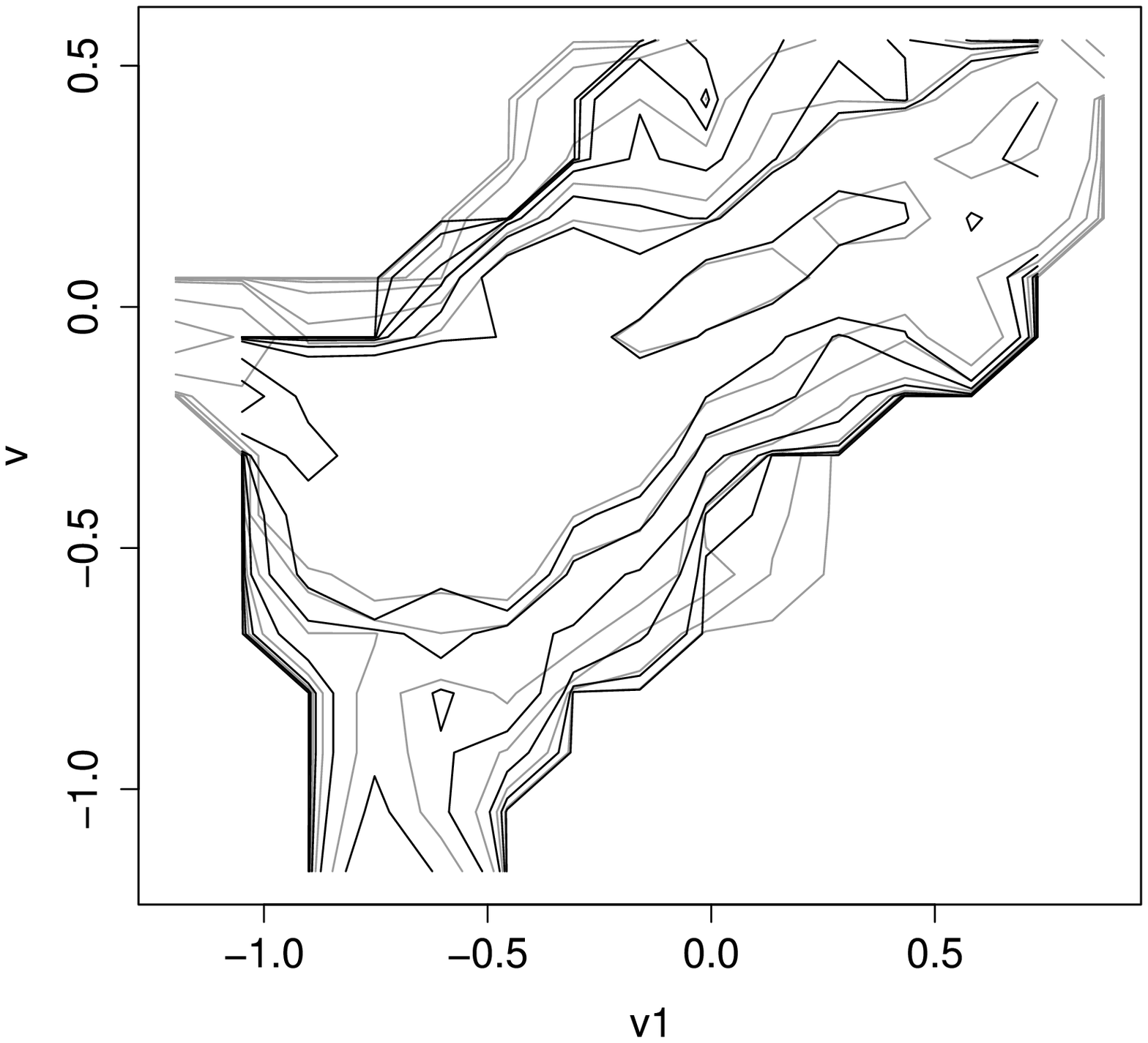}
 \put(-210,150){c)} &
\epsfxsize=2.78in
\epsffile{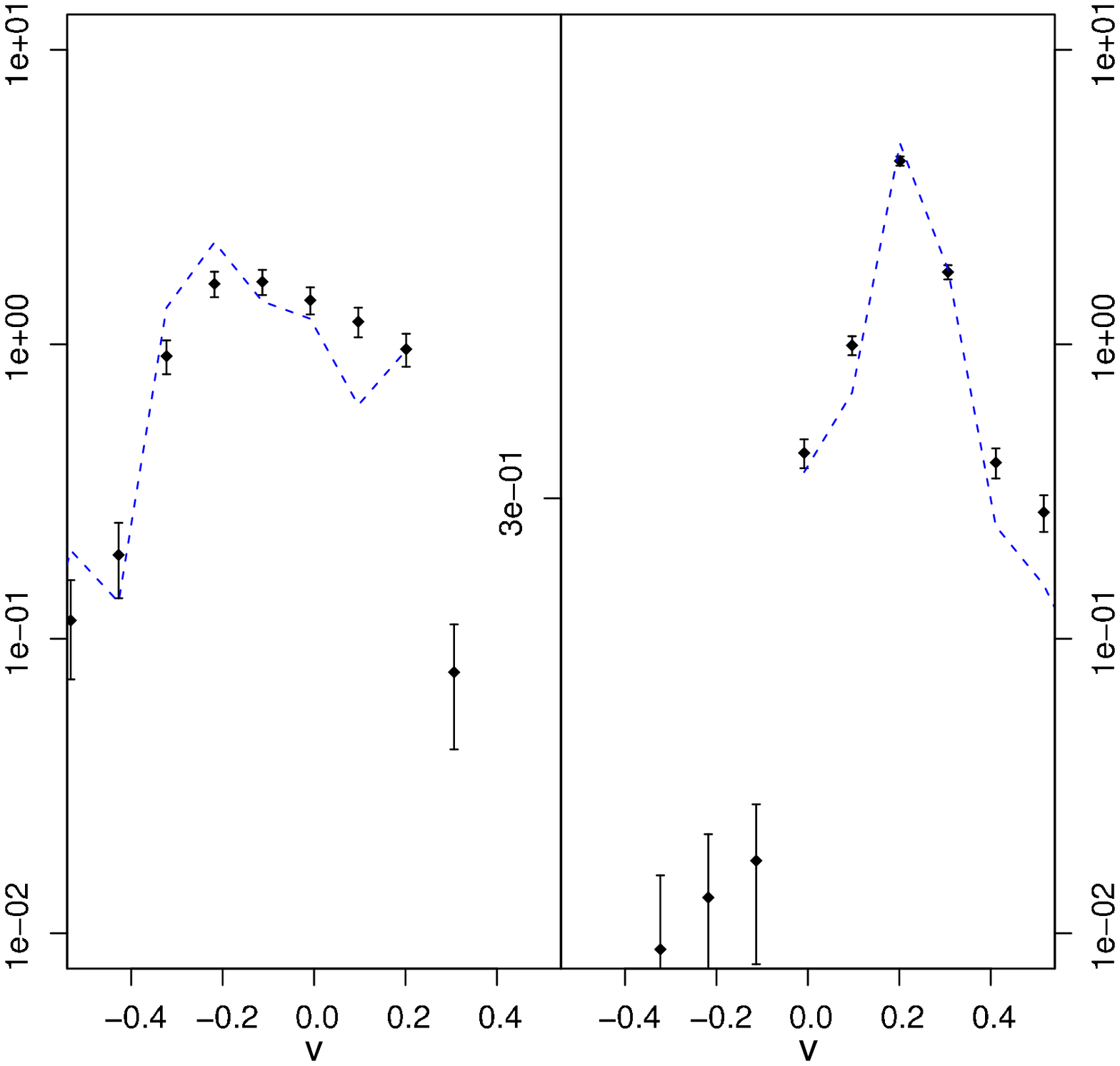}
 \put(-200,150){d)} \\ [-0.3cm]
\end{array}$
\caption{a) and c) give the contour plots of the pdfs of $p(u_{i},\tau_0|u_{i,1},\tau_1)$ (gray) and $p(u_{i},\tau_0|u_{i,1},\tau_1;u_{i,2},\tau_2)$ (black) for $\tau=0.15$ for u- and v-increments respectively. Slices at about $+/-0.8\sigma$ for u- and $+/-0.9\sigma$ for v-increments in b) and d) respectively show a high coherence between the two distributions, where the dotted line represents the data from $p(u_{i},\tau_0|u_{i,1},\tau_1;u_{i,2},\tau_2)$ and the errorbars are given by $\sqrt{N}$ with the number of events $N$ in the bin.}
\label{tau-cpdf-org}
\end{figure}
\end{center}

\begin{figure}[htb]
	\includegraphics[width=0.60\linewidth]{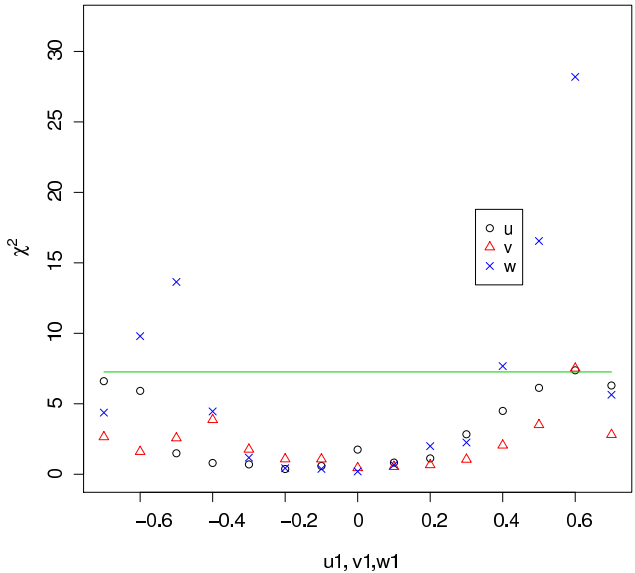}
        \caption{$\chi^2$ values for the conditioned pdfs with $\tau_0$ and $\Delta\tau=0.15$ against the $u_1$, $v_1$, and $w_1$-velocity increments bin. $\chi^2$ is calculated as the sum over all velocity bins $u$, $v$ and $w$. The line marks the value under which the probability of the distribution to be non-Markovian is less then 5\% in the classical $\chi^2$ theory.}
	\label{chi2-av-re}
\end{figure}

\begin{center}
\begin{figure}[htb]
$\begin{array}{c@{\hspace{0.3in}}c}  
 \multicolumn{1}{l}{\mbox{\bf }}  &
 \multicolumn{1}{l}{\mbox{\bf }} \\ [-0.5cm]
\epsfxsize=2.78in
\epsffile{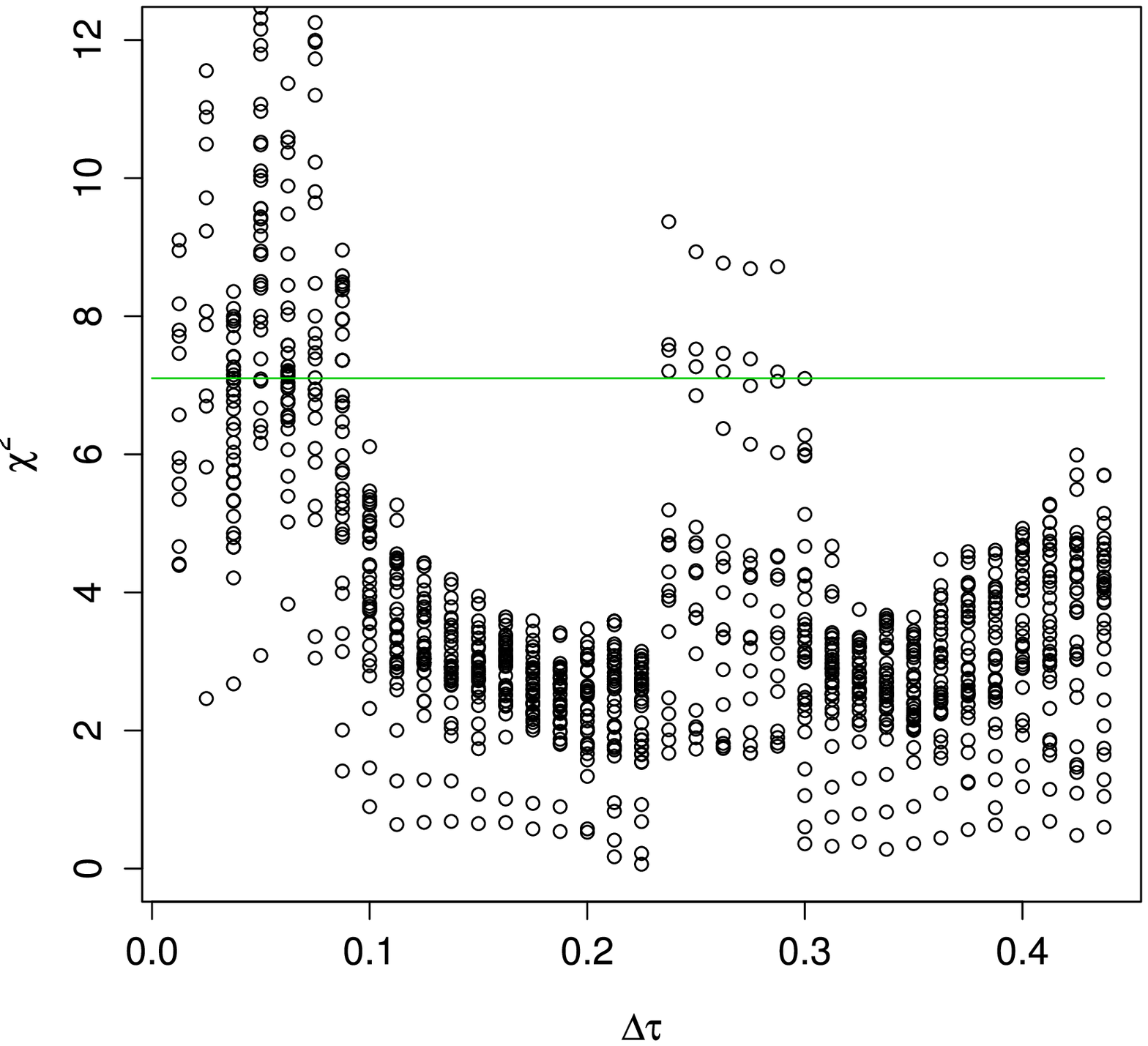}
 \put(-210,150){a)}  & 
\epsfxsize=2.78in
\epsffile{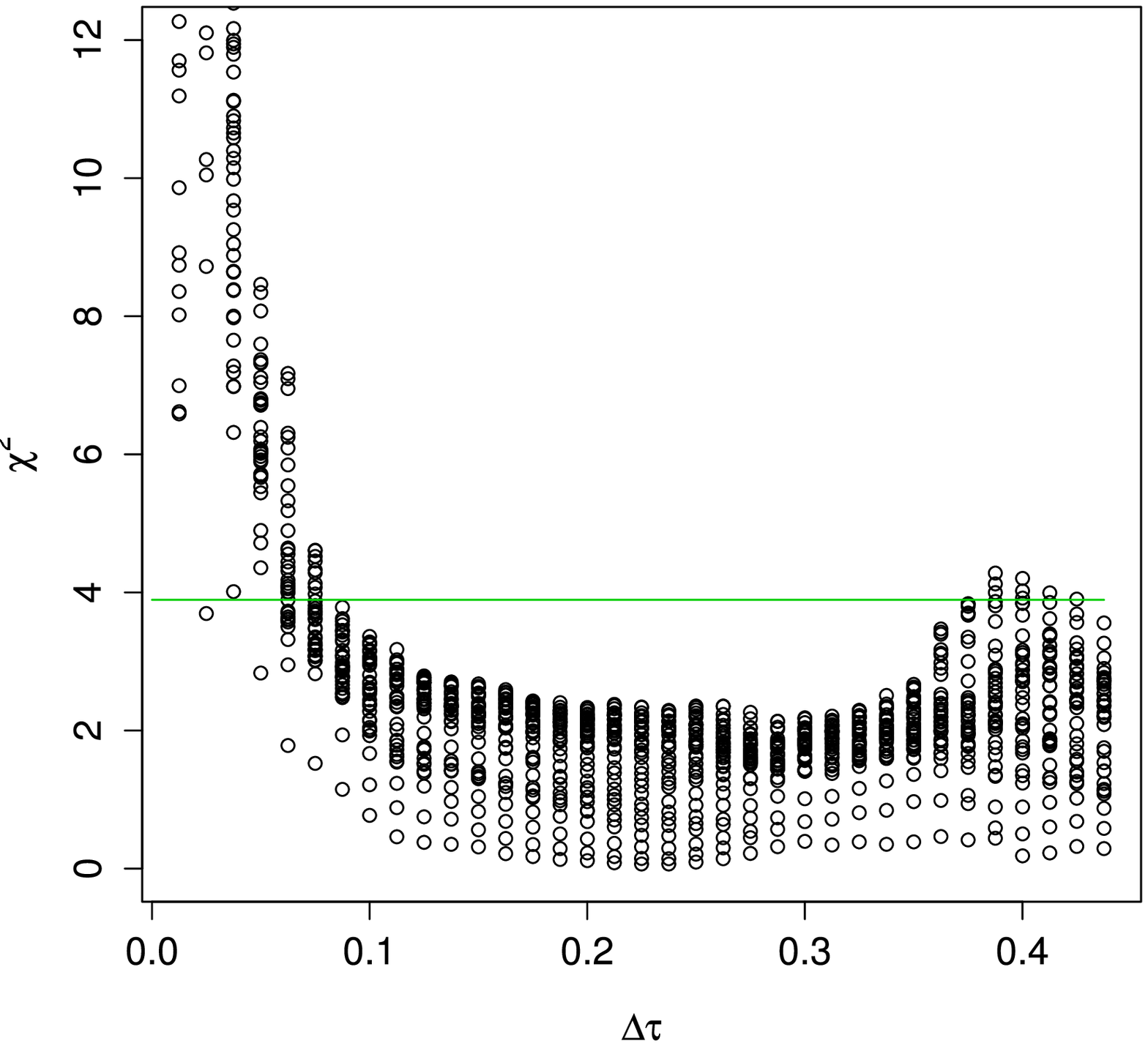}
\put(-200,150){b)}\\ [1.0cm]
\multicolumn{1}{l}{\mbox{\bf }}  &
\multicolumn{1}{l}{\mbox{\bf }}   \\ [-1cm]
\epsfxsize=2.78in
\epsffile{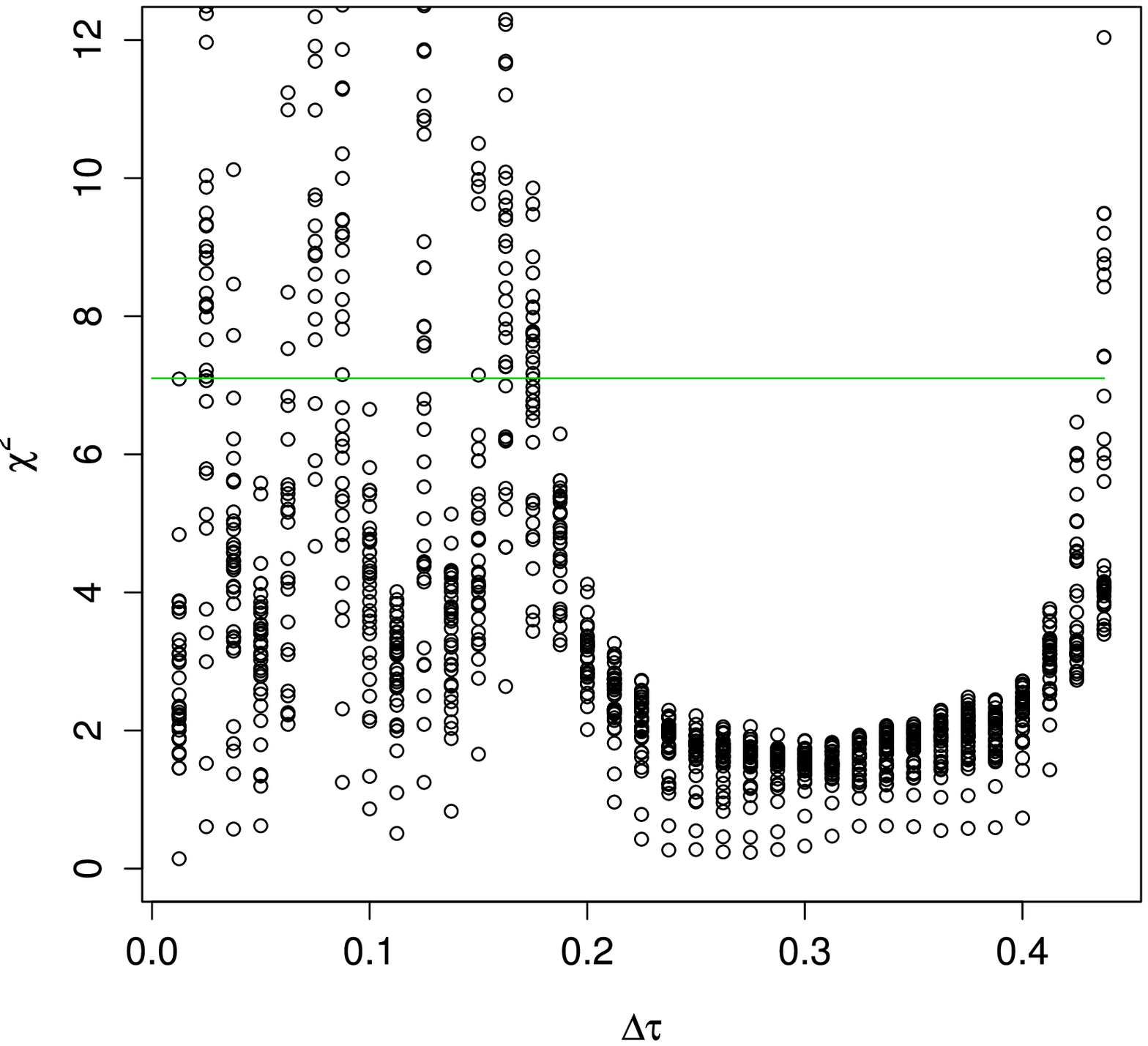}
 \put(-210,150){c)} &
\epsfxsize=2.78in
\epsffile{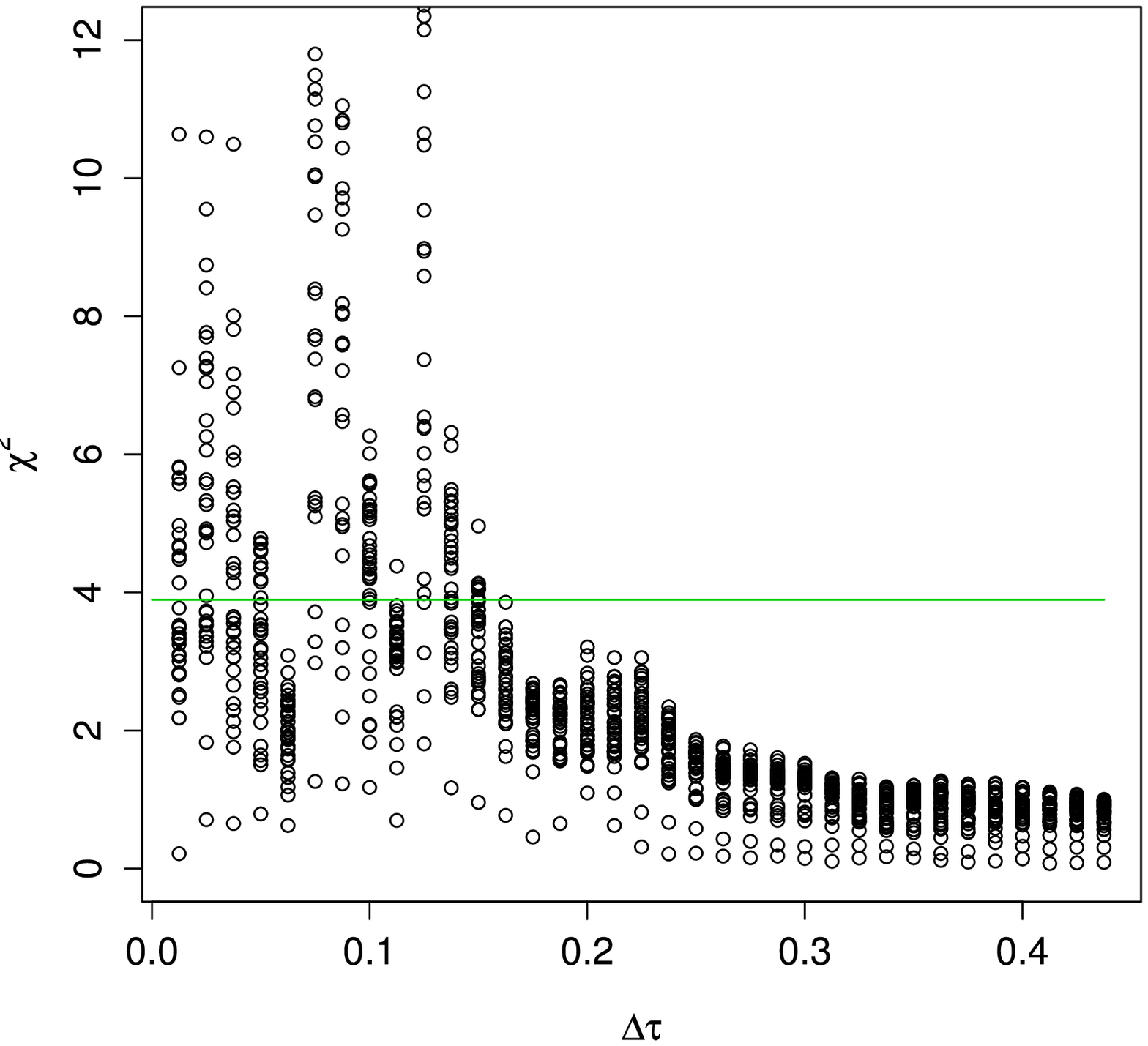}
\put(-200,150){d)} \\ [-0.3cm]
\end{array}$
        \caption{$\chi^2$ in dependency of $\Delta\tau$ for all bins a) and c) on the left and for only 10 well populated bins on the right b) and d), as a) and b) depict the u-component and c) and d) the v-component. It becomes evident, that Markov properties can be assumed in case of the smaller absolute value of the increments where enough data is available. The assumed validity of the Markov properties in the date here is $\Delta \tau \geq 0.15$ for the v-component and $0.075 \leq \Delta \tau \leq 0.375$ for the u-component.}
	\label{chi2-tau}
\end{figure}
\end{center}

\begin{center}
\begin{figure}[htb]
$\begin{array}{c@{\hspace{0.3in}}c}  
 \multicolumn{1}{l}{\mbox{\bf }}  &
 \multicolumn{1}{l}{\mbox{\bf }} \\ [-1.5cm]
\epsfxsize=2.78in
\epsffile{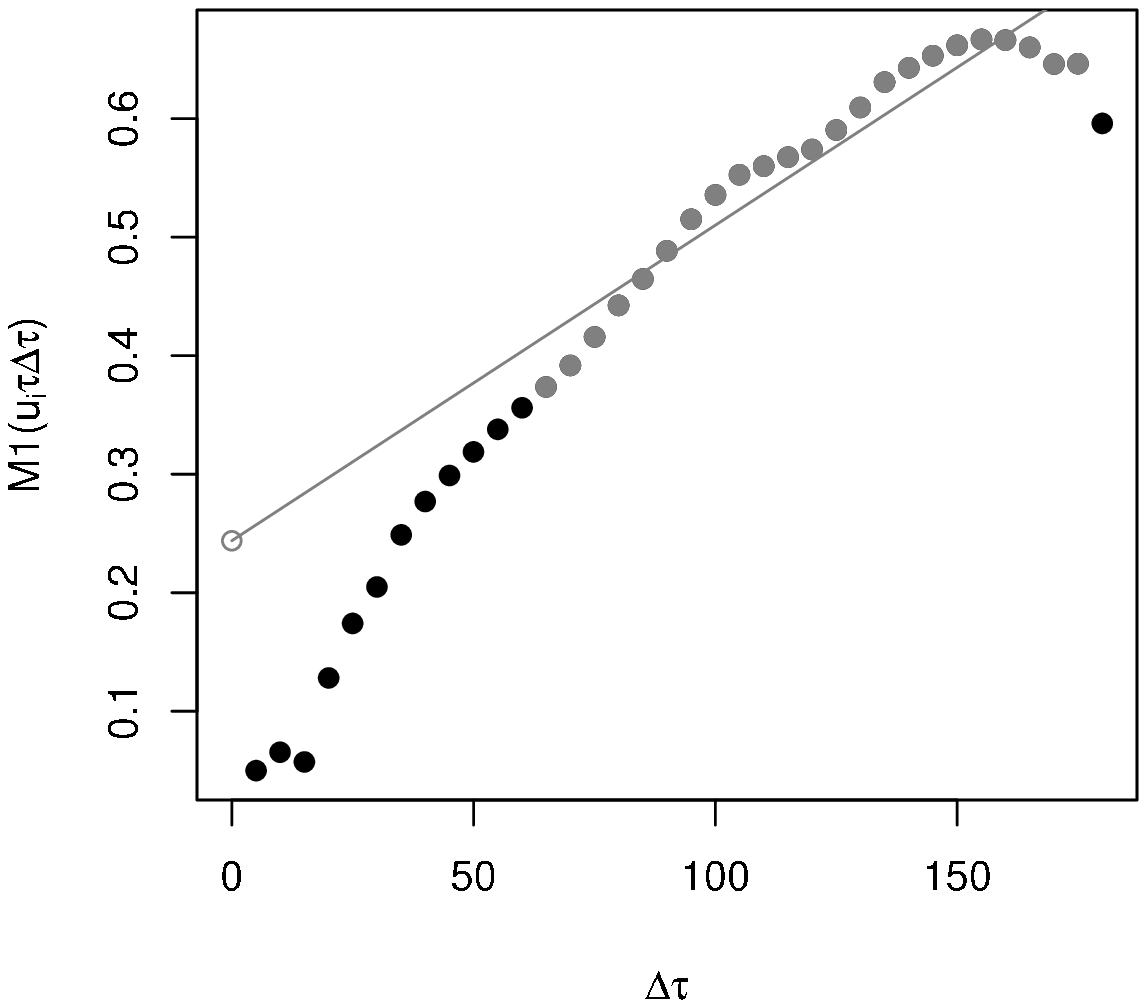}
 \put(-210,150){a)}  & 
\epsfxsize=2.78in
\epsffile{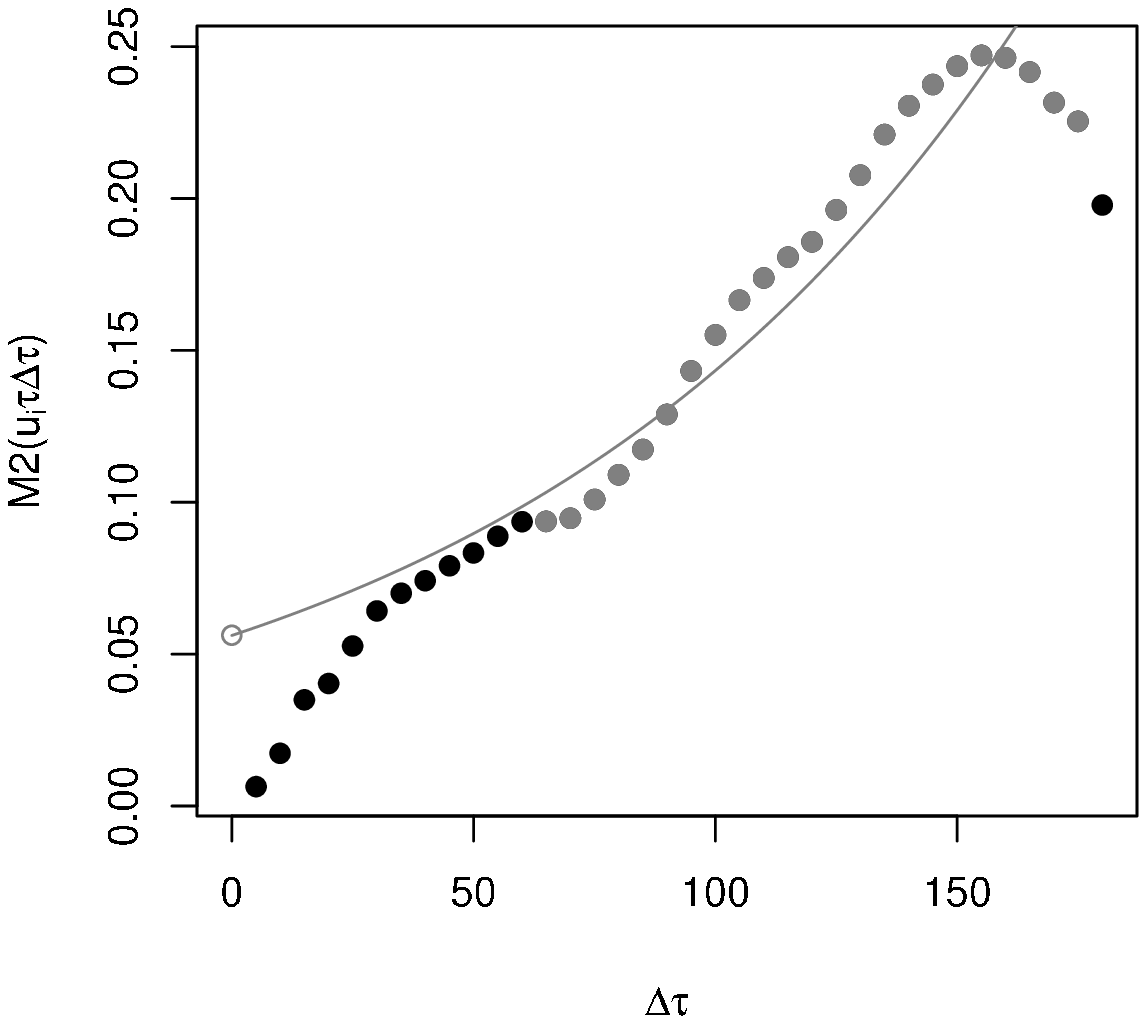}
\put(-200,150){b)}\\ [1.0cm]
\multicolumn{1}{l}{\mbox{\bf }}  &
\multicolumn{1}{l}{\mbox{\bf }}   \\ [-0.9cm]
\epsfxsize=2.78in
\epsffile{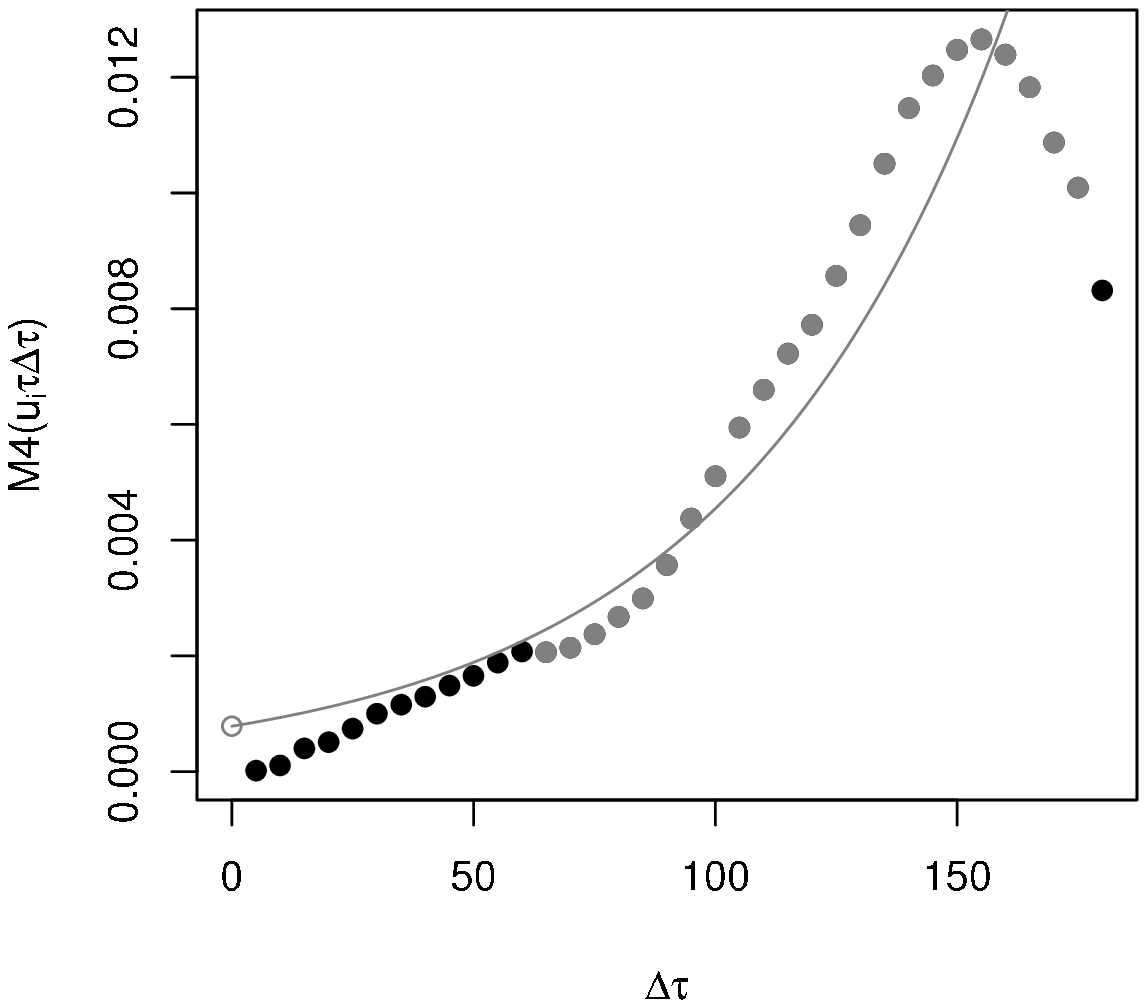}
 \put(-210,150){c)} &
\epsfxsize=2.78in
\epsffile{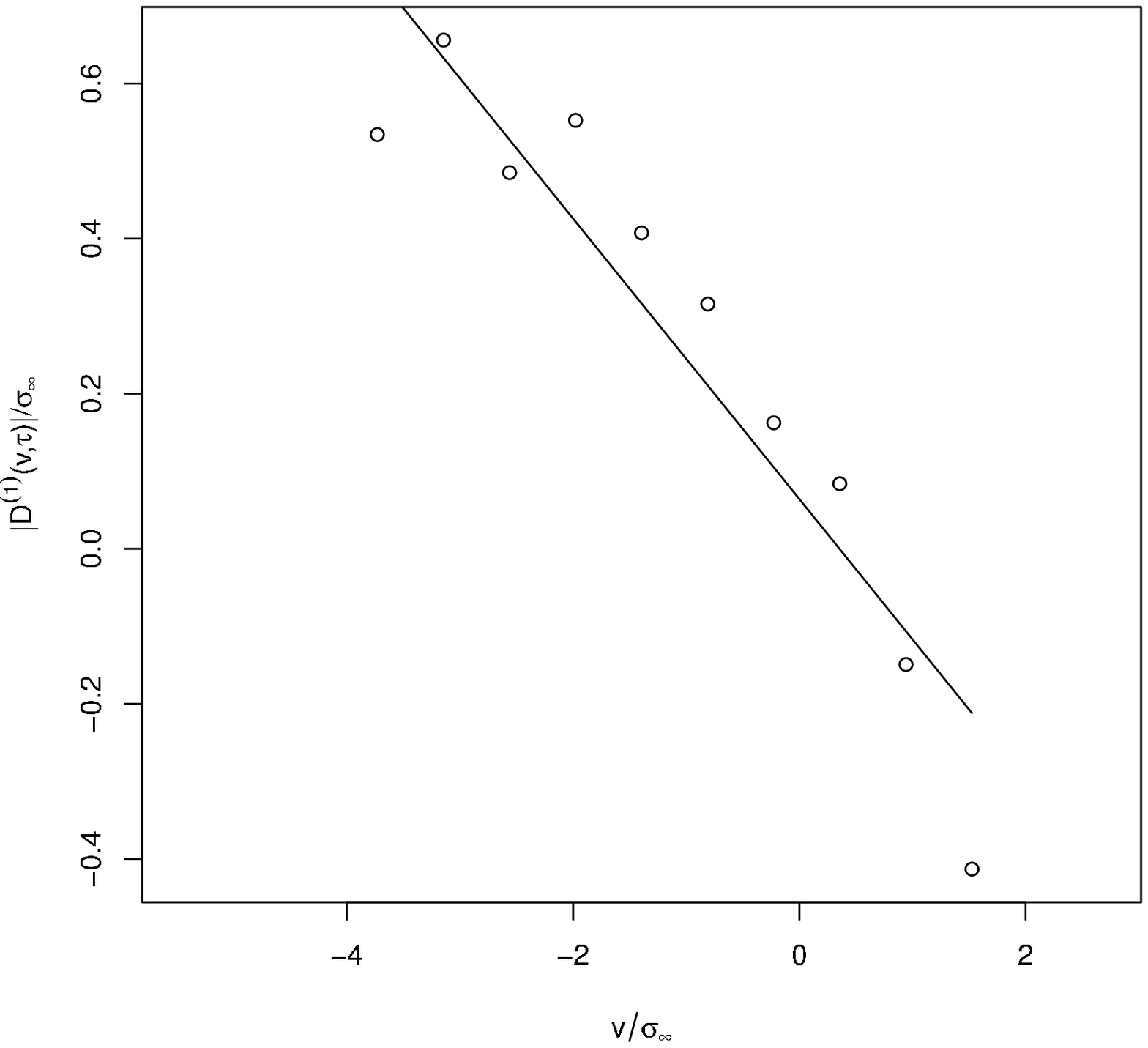}
 \put(-200,150){d)} \\ [-0.3cm]
\multicolumn{1}{l}{\mbox{\bf }}  &
\multicolumn{1}{l}{\mbox{\bf }}   \\ [-0.0cm]
\epsfxsize=2.78in
\epsffile{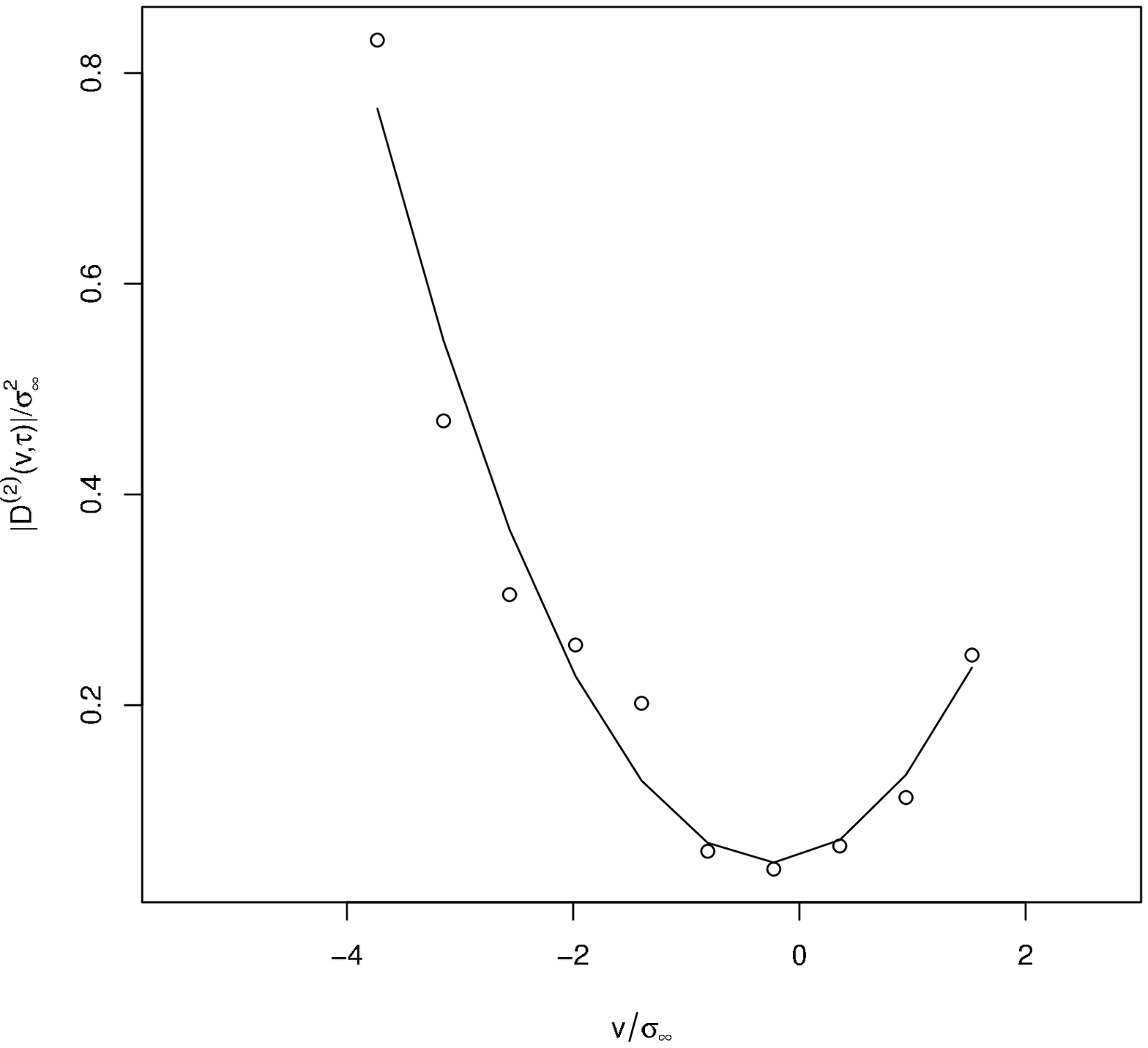}
 \put(-210,150){e)} &
\epsfxsize=2.78in
\epsffile{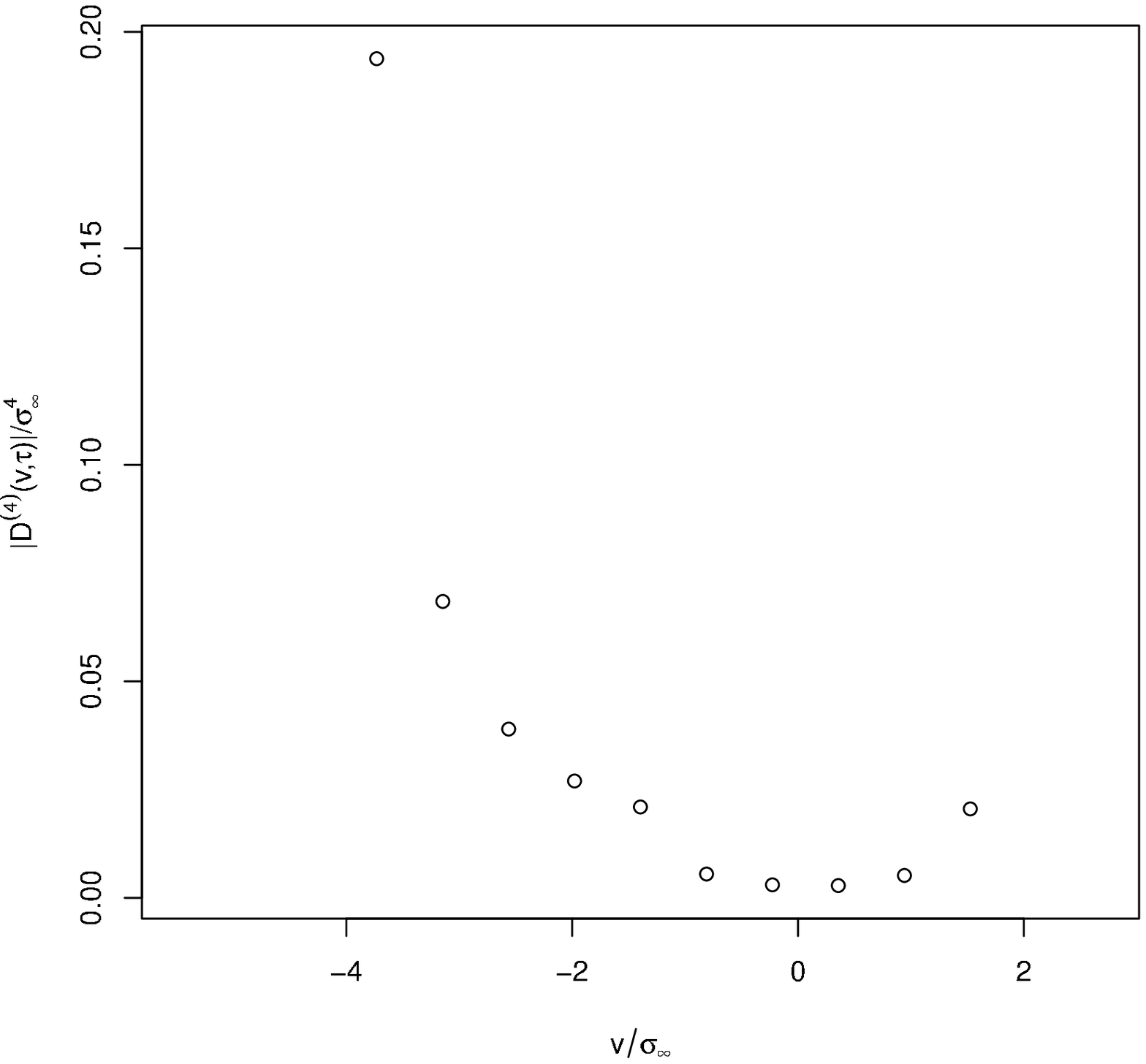}
 \put(-200,150){f)} \\ [-0.3cm]
\end{array}$
        \caption{Examples of the fits of $M^{(1)}$, $M^{(2)}$ and $M^{(4)}$ in fig. a), b) and c) respectively for $\tau=0.15$. The used range is plotted in gray, while black dots have been excluded. The resulting normed $D^{(1)}$, $D^{(2)}$ and $D^{(4)}$  plotted on all used bins with fits for $D^{(1)}$ and $D^{(2)}$ in d), e) and f).}
	\label{fit-ds}
\end{figure}
\end{center}

The first analysis of the time series has been conducted using the dataset of the point marked with the number 218 as a reference in time scales $\tau$. We are later interested to use the Taylor hypothesis. The mean flow field direction is oriented in the x-y-plane. Therefore we focussed here on the analysis of the u- and v-velocity increments. However, the complete analysis has also been done for the spanwise direction, even though for brevity these results might not all be presented here.\\
To give a first idea of the distributions of the velocity increments, the non conditioned histograms of the single components are shown in fig.\ref{pdf-comp} for $\tau_0=0.15$. It is obvious, that for the u and v-components there is some skewness in the distribution and all distributions are non-Gaussian.\\
Next we discuss at which scales the flow field does show Markov properties. For all geometrical points a dataset of a time series of 10922 points in time was recorded. Since this was very short, only 15 bins have been used for the evaluation.\\
Fig.\ref{tau-cpdf-org} shows the contours of the conditioned probability density function for $p(u_{i},\tau_0|u_{i,1},\tau_1)$ and $p(u_{i},\tau_0|u_{i,1},\tau_1;u_{i,2},\tau_2)$ for $\Delta\tau=\tau_1-\tau_0=0.15$, $\tau_0=0.15$ and $\tau_2=\tau_0+2\Delta\tau$ for the velocity components u and v. Also slices of the contour-plot are given to give an impression of the shape of the conditioned pdfs. The shapes of the pdfs do not appear very smooth as it has to be kept in mind, that the over all number of data is just above 10000 and the plots are logarithmic. However, the similarity of the two contours suggest the existence of Markov properties.\\
Next, a more quantifying test is done to check if Markov properties are given. To evaluate the Markov properties, now a $\chi^2$-test has been performed for the distributions of $p(u_{i},\tau_0|u_{i,1},\tau_1)$ and $p_2(u_i,\tau|u_{i,0},\tau_0;u_{i,1},\tau_1)$ (see \cite{bronstein91})
\begin{eqnarray}
\chi^2=\sum_i \frac{\left(p_1(i)-p_2(i)\right)^2}{p_1(i)}.
\label{chi2}
\end{eqnarray}
Here $p_1(i)$ represents the first probability density - in this first case $p(u_{i},\tau_0|u_{i,1},\tau_1)$, $i$ is the bin number and $p_2(i)$ is the second probability density respectively - here $p(u_{i},\tau_0|u_{i,1},\tau_1;u_{i,2},\tau_2)$. The test quantifies the difference between two pdfs from data with errors. Usually as $p_1$ a theoretically know pdf is taken. For our purpose of the estimation of the existence of Markov properties, this might lead to cases where $p_1(i)=0$. In such cases we set $\chi^2$ has been set to $0$. Thus $\chi^2=0$ events had to be evaluated separately.\\x
15 bins have been used for evaluation, giving 15 degrees of freedom. Taking the average of the $\chi^2$ values for all the evaluated distributions shown in fig.\ref{chi2-av-re} for all velocity components, we obtain an impression of the validity of the assumption of Markov properties.\\
In fig.\ref{chi2-av-re} it becomes evident, that the bins containing few incidences further from the mean values, show stronger deviances in the distributions. This might be due to statistical reasons which lead to very low binning values, however it might also be caused by a non-Markovian behavior of the stronger fluctuations in the flow at some scales. This could only be clarified if the dataset would be increased for a longer simulation period.\\
Due to this insecurity in the Markov analysis, five bins have been excluded from the evaluation for the further analysis: The three outer bins to the lowest negative velocity increments and the two bins containing the highest velocity increments. Fig.\ref{chi2-tau} shows the dependency of the average $\chi^2$ on $\Delta\tau$ for $0.0125 \leq \Delta \tau \leq 0.45$ for the velocity components, a) and c) including all bins and in b) and d) using only 10 bins after the exclusion of the five bins containing to few data. Note, that with the decrease of degrees of freedom, the limit for the criterion of the probability being $\leq 5\%$ for non-Markov properties also decreases.
Using the 10 bins we can expect Markov properties at $\Delta \tau \geq 0.15$ for the v- and w-component while for the u-component we supposed to have a Markovian field for $0.075 \leq \Delta \tau \leq 0.375$. Applying the Taylor hypothesis (eq. (\ref{Taylor})), this would correspond to a so called Markov length of $\lambda_M= 0.011$ for the u-velocity and $\lambda_M=0.018$ for the v-velocity respectively. Note the comparable values to $\lambda$ in table \ref{spacing} as proposed by \cite{lueck2006}.\\
Within the estimated validity range for the Markow properties, now the Kramers-Moyal coefficients have been determined for each $\tau$ and $\Delta\tau$. Therefore for each $\tau$ a fit for $lim_{\Delta \tau \rightarrow 0}$ corresponding to equation (\ref{M-coeff}) has been done. To estimate $D^{(1)}$ a linear fit was used, whereas for $D^{(2)}$ and $D^{(4)}$ logarithmic fits have been used, since they fitted to course of the curve better and avoided unphysical negative values. As an example in fig.\ref{fit-ds} a), b) and c) the fit for the normed v-velocity increments at $\tau=0.15$ is shown for $D^{(1)}$, $D^{(2)}$ and  $D^{(4)}$.\\
The $D^{(1)}(u_i,\tau)$ and $D^{(2)}(u_i,\tau)$ have further been fitted by a linear and a quadratic function over all $u_{i,\tau}$, as described by Renner in \cite{renner01} for isotropic turbulence (see fig.\ref{fit-ds} d) and e)). These fits were in the following used for the reconstruction.\\
Even though we cannot assume $D^{(4)}(u_i,\tau) = 0$ in the dataset used, since $D^{(4)}<<D^{(2)}$ we take the Fokker-Planck equation (\ref{FP}) as an approximation.\\

\subsection{Reconstruction in time}
\begin{center}
\begin{figure}[htbp]
$\begin{array}{c@{\hspace{0.3in}}c}  
 \multicolumn{1}{l}{\mbox{\bf }}  &
 \multicolumn{1}{l}{\mbox{\bf }} \\ [-0.5cm]
\epsfxsize=2.78in
\epsffile{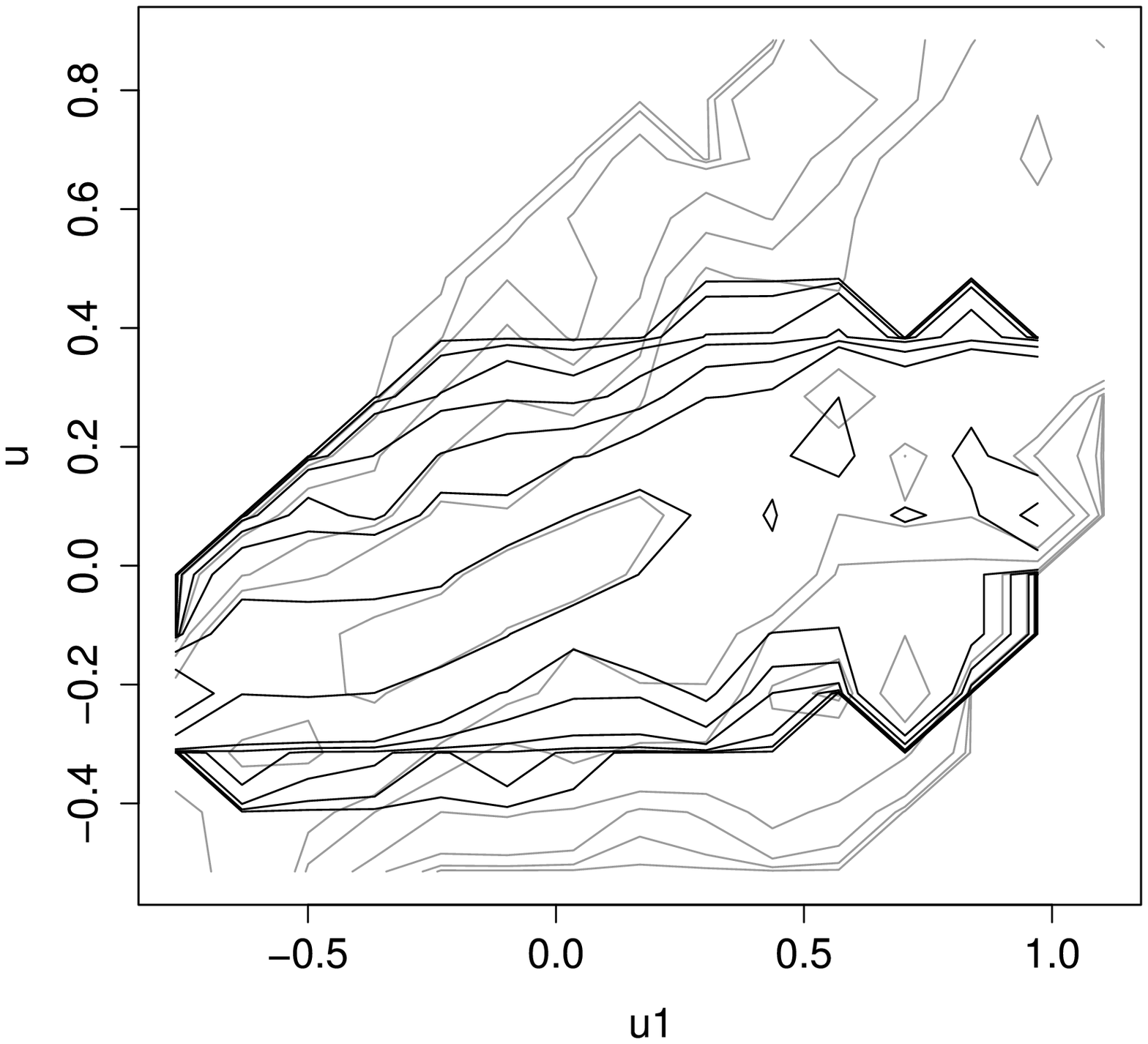}
 \put(-210,150){a)}  & 
\epsfxsize=2.78in
\epsffile{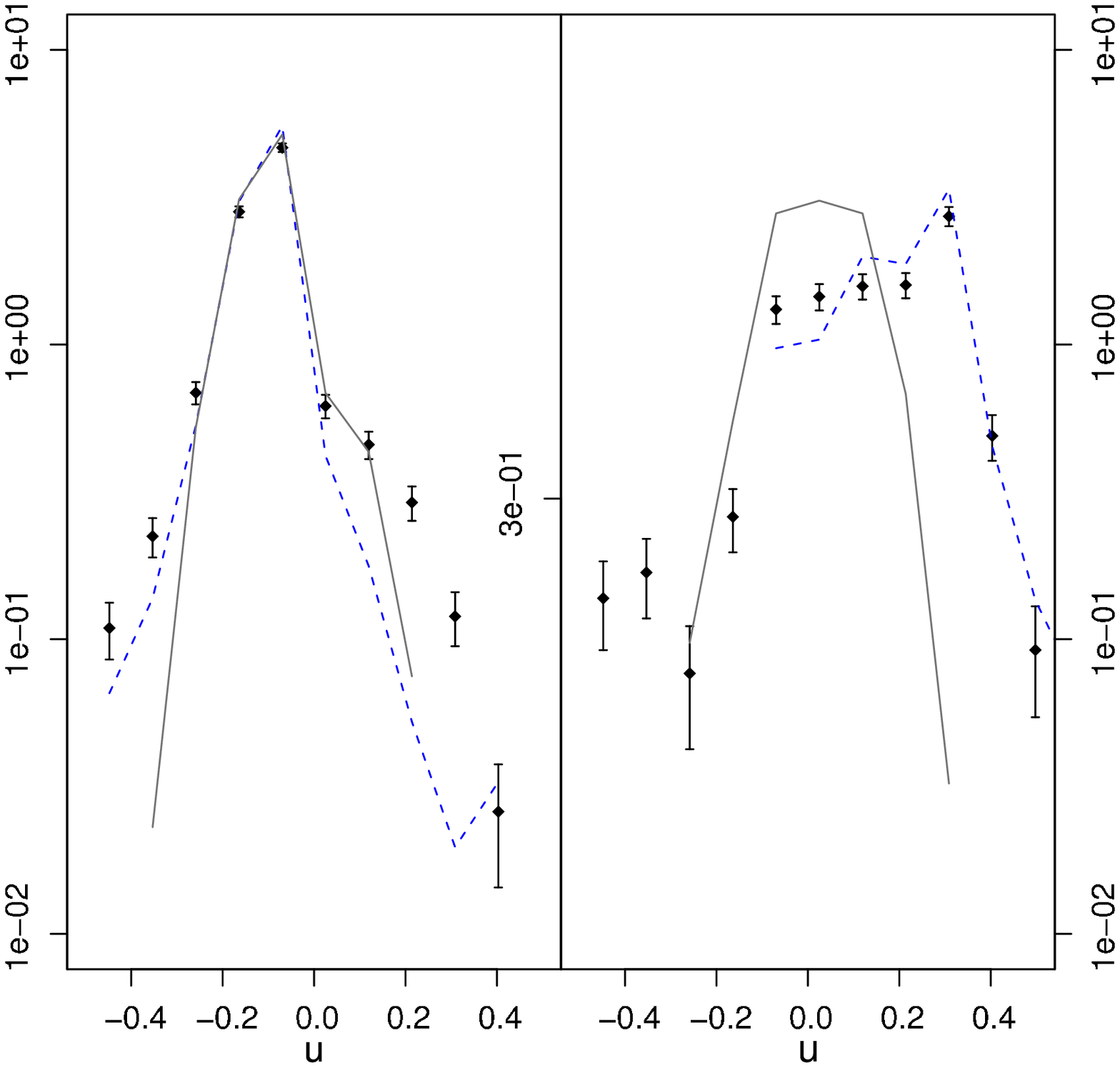}
\put(-200,150){b)}\\ [1.0cm]
\multicolumn{1}{l}{\mbox{\bf }}  &
\multicolumn{1}{l}{\mbox{\bf }}   \\ [-1cm]
\epsfxsize=2.78in
\epsffile{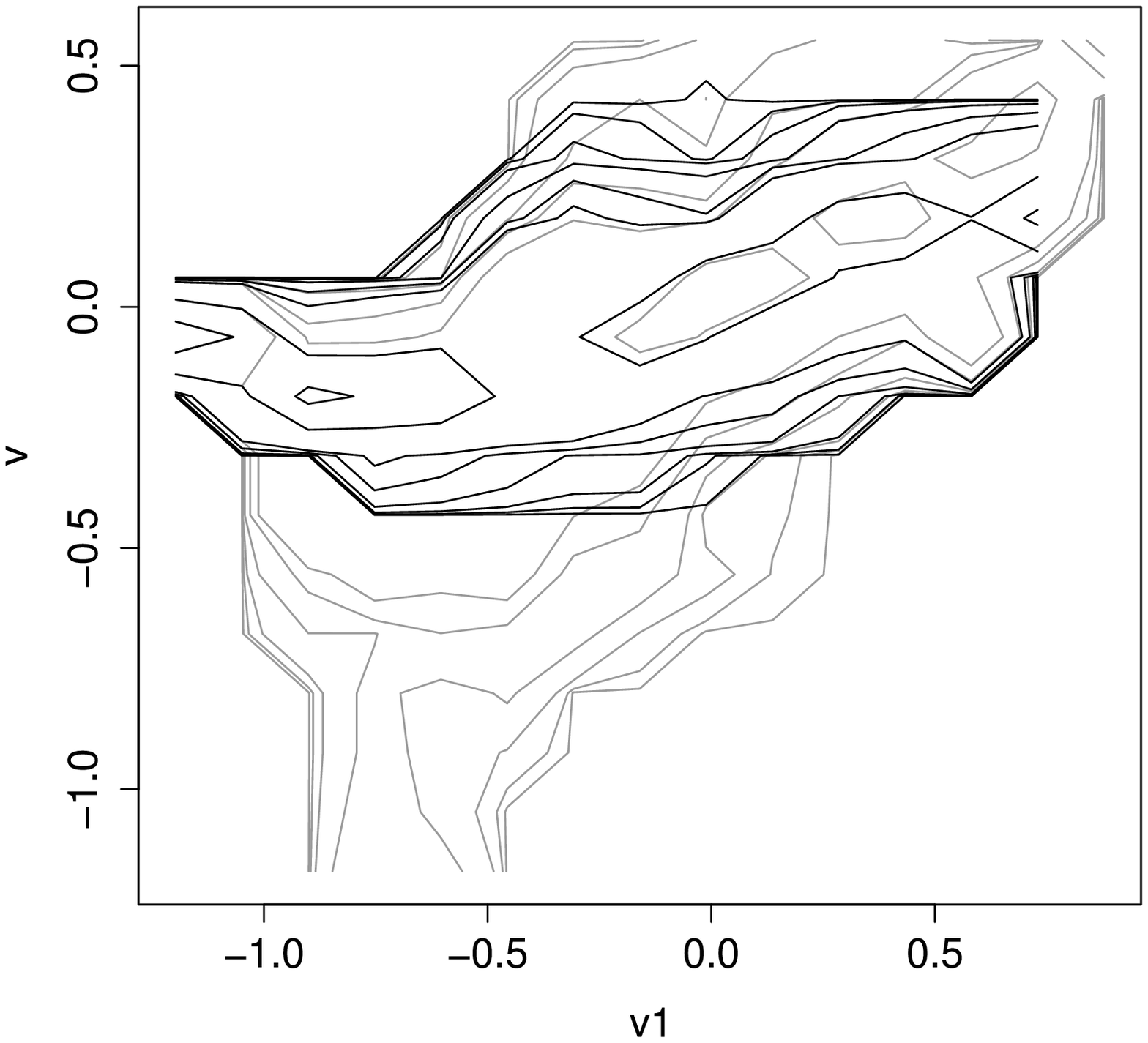}
 \put(-210,150){c)} &
\epsfxsize=2.78in
\epsffile{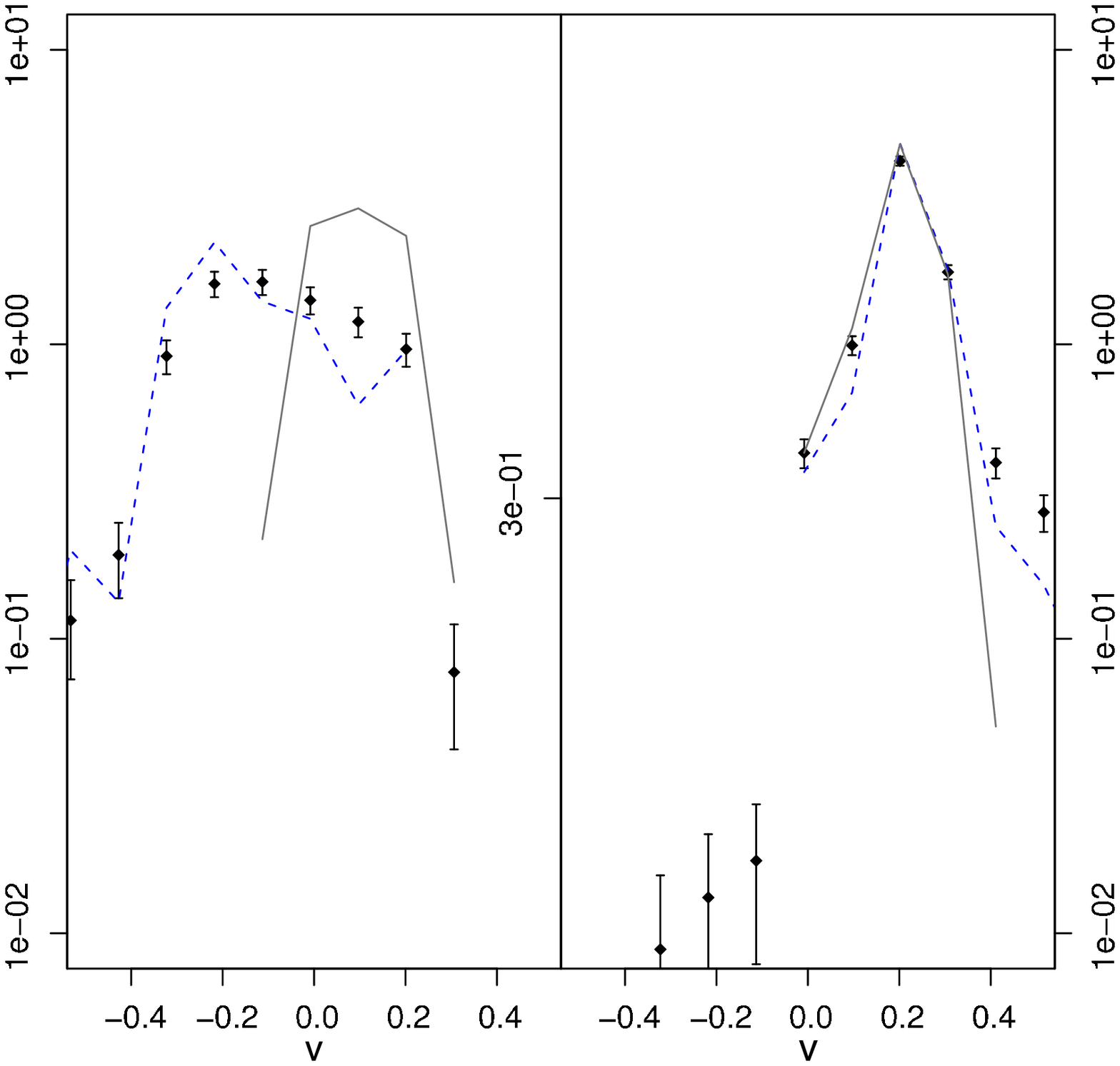}
 \put(-200,150){d)} \\ [-0.3cm]
\end{array}$
\caption{Comparison of the contour plots in a) and c) of the pdfs of$p(u_{i},\tau_0|u_{i,1},\tau_1)$ and $p(u_{i},\tau_0|u_{i,1},\tau_1;u_{i,rec},\tau_2)$ for $\tau=0.15$ in case of the reconstructed time series $u_{i,rec}$ for u- and v-increments. Again slices at about $+/-0.8\sigma$ for u and $+/-0.9\sigma$ for v in b) and d) respectively give an impression of the distributions. Here the dotted line represents the data from $p(u_{i},\tau_0|u_{i,1},\tau_1;u_{i,2},\tau_2)$ and the solid line the data from $p(u_{i},\tau_0|u_{i,1},\tau_1;u_{i,rec},\tau_2)$.}
\label{time-rec}
\end{figure}
\end{center}

\begin{center}
\begin{figure}[htbp]
$\begin{array}{c@{\hspace{0.3in}}c}  
 \multicolumn{1}{l}{\mbox{\bf }}  &
 \multicolumn{1}{l}{\mbox{\bf }} \\ [-0.5cm]
\epsfxsize=2.78in
\epsffile{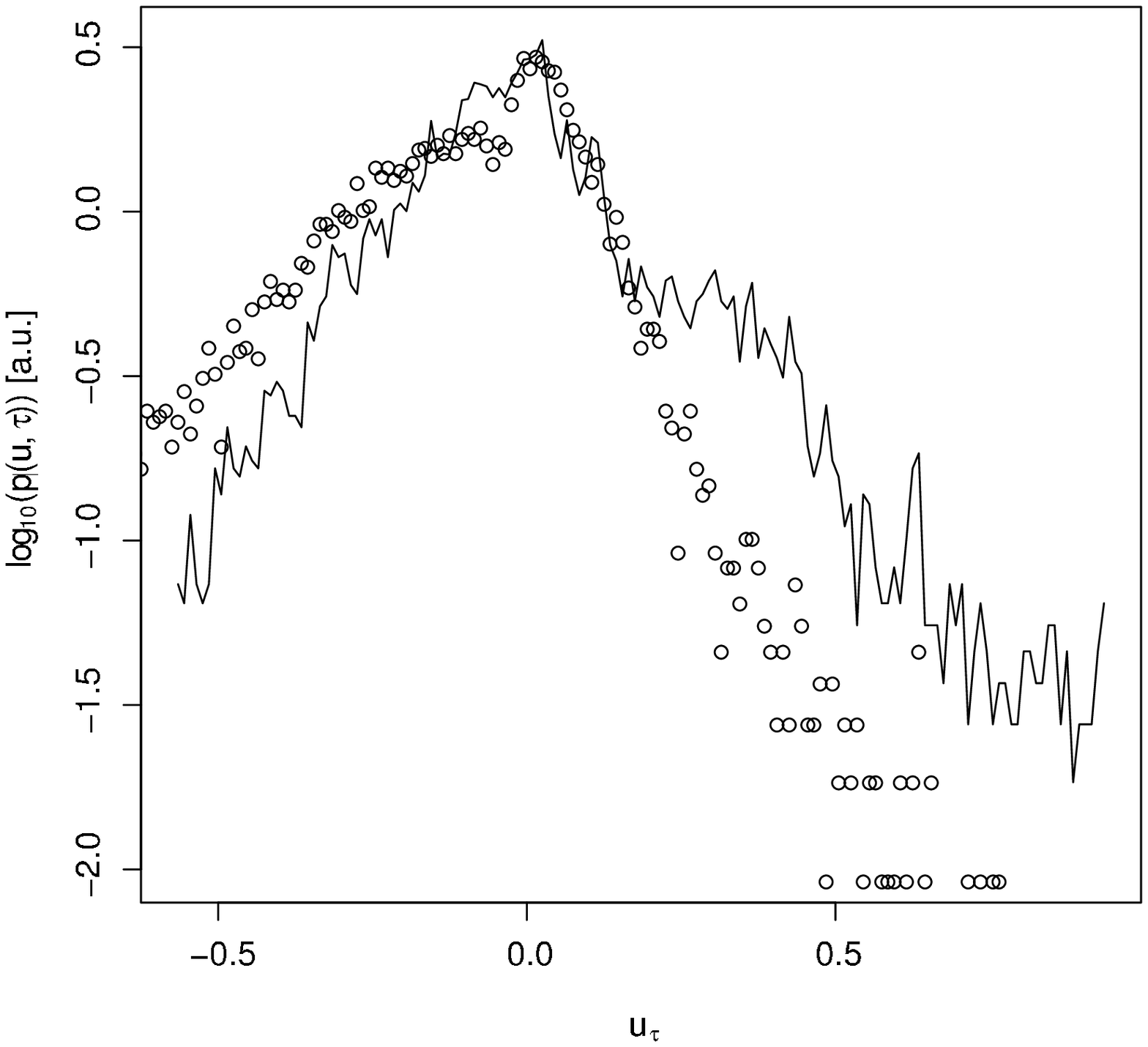}
 \put(-210,150){a)}  & 
\epsfxsize=2.78in
\epsffile{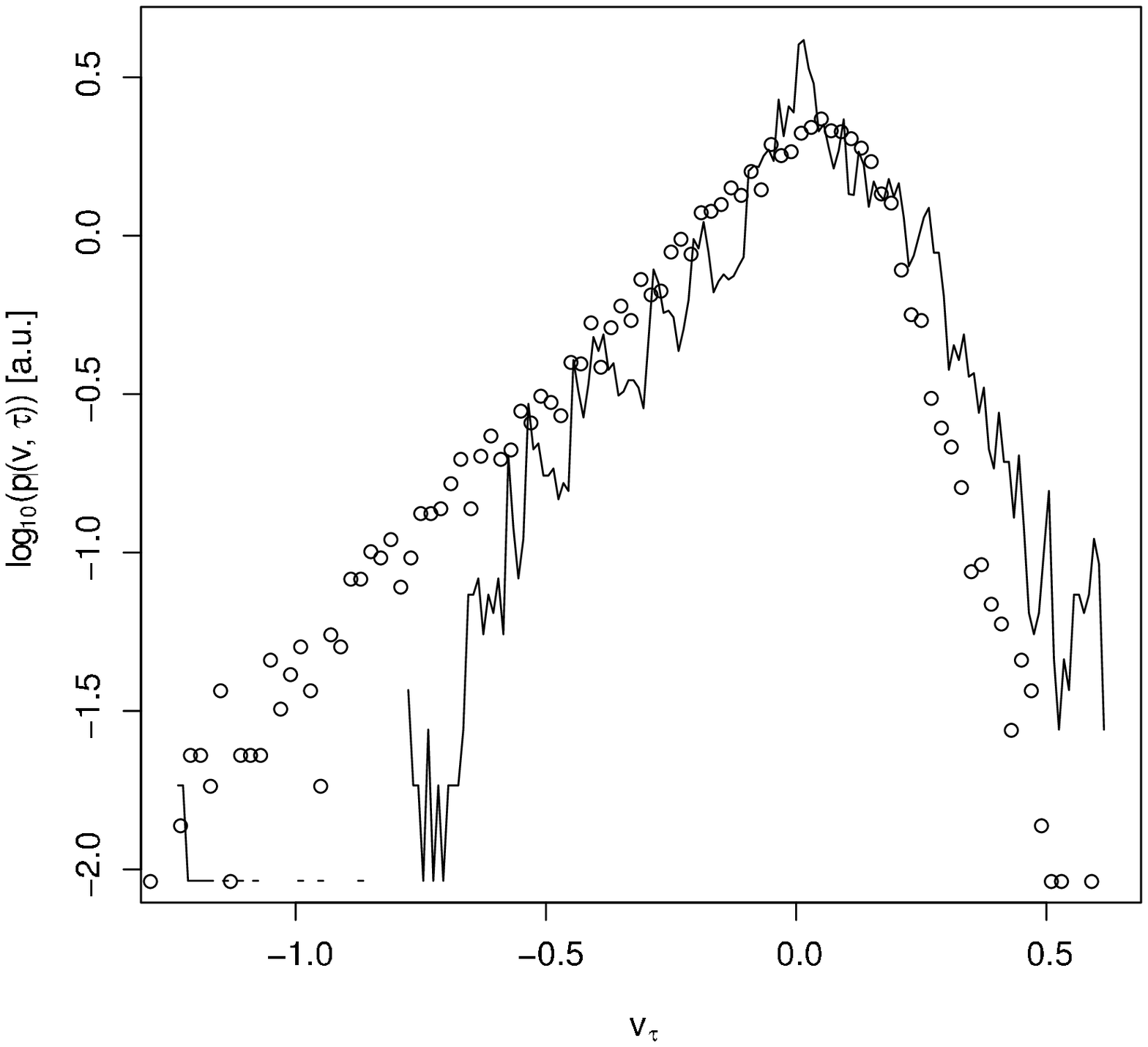}
\put(-200,150){b)}\\ [1.0cm]
\end{array}$
\caption{Logarithmic plot of the non conditioned pdfs $p(u,\tau_0)$ (line) and $p(u_{rec},\tau_0)$ dots at $\tau_0=\tau_1=0.15$ showing good general coincidence and some differences in particular due to the different skewness of the distributions.}
\label{pdf-rec}
\end{figure}
\end{center}

To get an estimation of the accuracy of a reconstruction assuming a Gaussian field, the reconstruction has been done using only the first two Kramers-Moyal coefficients.\\
The obtained $D^{(1)}(u_i,\tau)$ and $D^{(2)}(u_i,\tau)$ are used to reconstruct the time series with a Langevin equation (\ref{Langevin}) modified for time scales. In fig.\ref{time-rec} the pdfs $p(u_{i},\tau_0|u_{i,1},\tau_1)$ and $p(u_{i},\tau_0|u_{i,1},\tau_1;u_{i,rec},\tau_2)$ of the u- and v-velocity increments for $\tau_0=0.15$ and $\tau_1=0.30$ are compared. Here $u_{i,rec}(\tau_2)$ is the reconstructed data at a $3\tau$ scale. Below again slices at about $+/-0.8\sigma$ for u and $+/-0.9\sigma$ are shown.\\
The pdfs of the increments show a non perfect, but quite good reconstruction. Fig. \ref{pdf-rec} shows the semi-logarithmic plot of the pdf of the original data against the reconstructed data for $\tau_0=0.15$. Although certainly the dataset was still short, the general shape of the curve is met in most cases (see table \ref{rec-data}). The main reason for deviations is most likely the approximation for the Kramers-Moyal expansion, since $D^{(4)}\not =0$ as the higher order Moments are not met perfectly.\\

\begin{table}[htb]
\begin{tabular}{|c|c|c|c|c|c|c|c|c|}
\hline
Comp&Type&Mean&$\sigma$&Skewness&Ex-Kurtosis\\
\hline
x&data&0.001&0.224&0.75&1.46\\
\hline
x&recon&-0.1246&0.243&-1.08&2.63\\
\hline
y&data&0.003&0.216&-0.74&2.06\\
\hline
y&recon&-0.089&0.258&-1.37&3.07\\
\hline
z&data&0.0008&0.155&0.26&2.03\\
\hline
z&recon&-0.051&0.158&-1.00&4.23\\
\hline
\end{tabular}
\label{rec-data}
\caption{Statistical properties of the distributions of the original data (marked with data) and the reconstructed time series (recon) for $\tau=0.15$. Notable is that the higher order moments are not reconstructed correctly.}
\end{table}
\begin{center}

\begin{figure}[htbp]
$\begin{array}{c@{\hspace{0.3in}}c}  
 \multicolumn{1}{l}{\mbox{\bf }}  &
 \multicolumn{1}{l}{\mbox{\bf }} \\ [-0.5cm]
\epsfxsize=2.78in
\epsffile{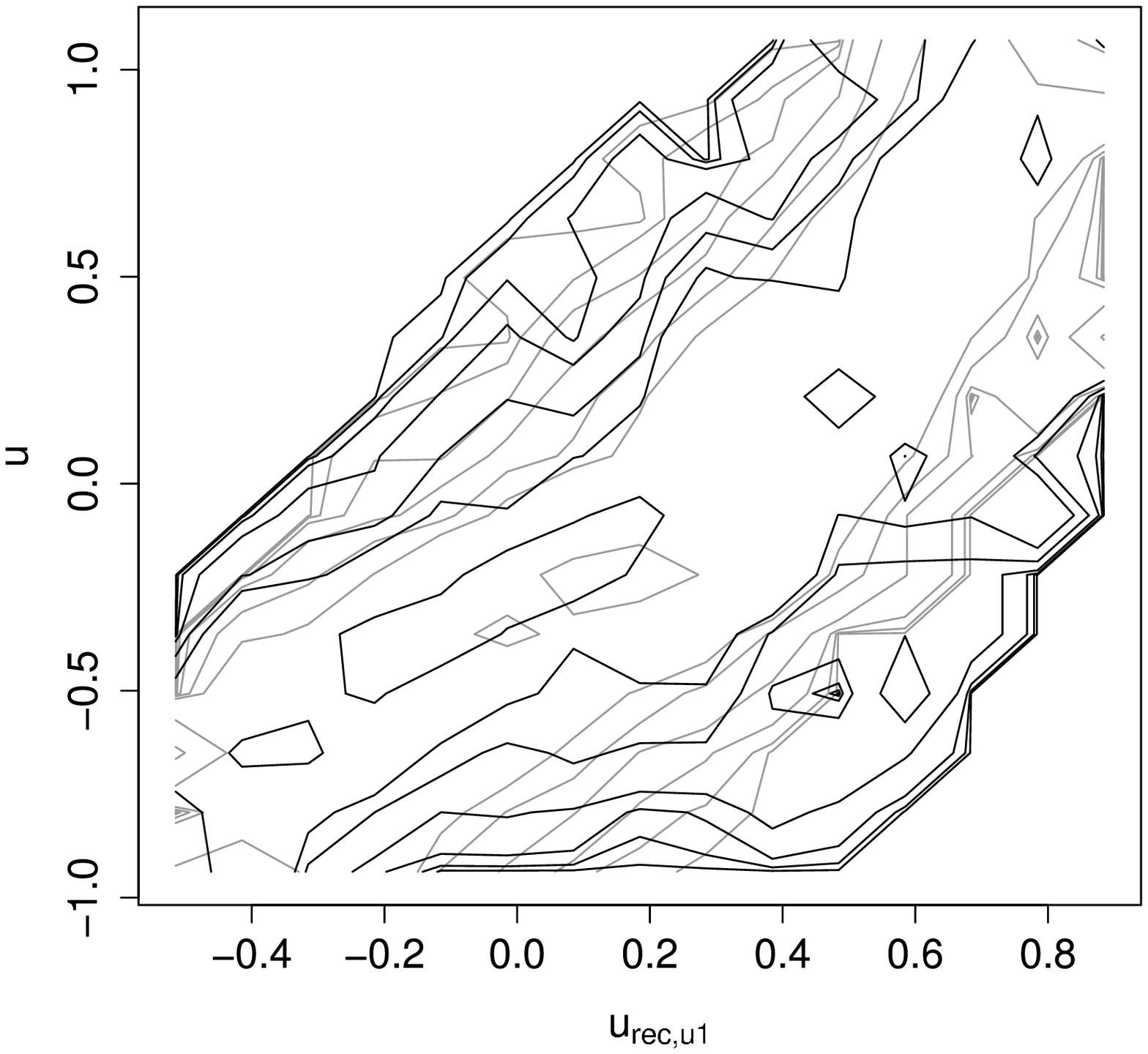}
 \put(-210,150){a)}  & 
\epsfxsize=2.78in
\epsffile{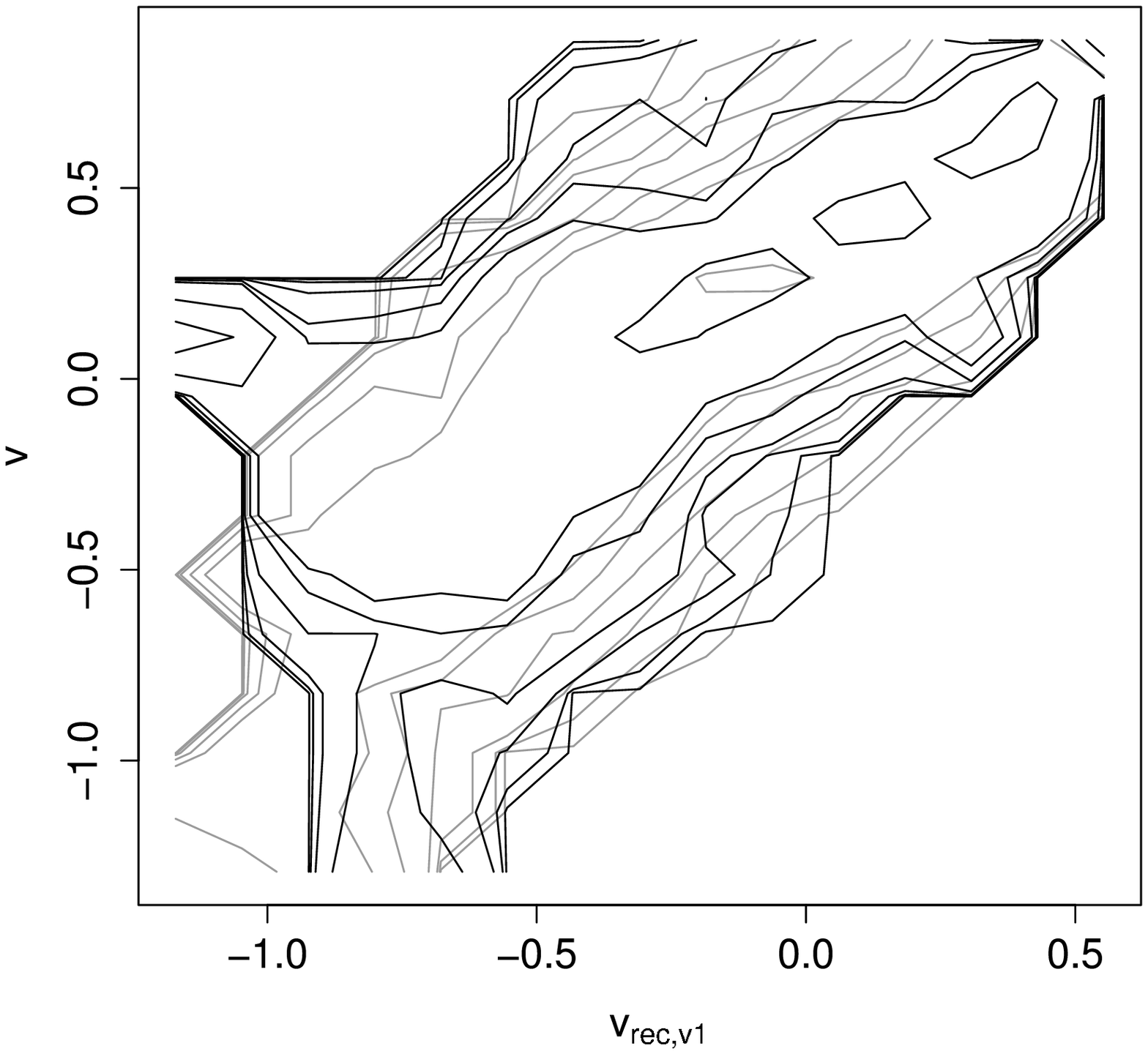}
\put(-200,150){b)}\\ [1.0cm]
\end{array}$
\caption{Contour plot of the single conditioned pdfs $p(u,\tau|u_{0},\tau_0)$ (black) and $p(u,\tau|u_{rec},\tau_0)$ (grey) at $\tau=\tau_0=0.15$ showing the coincidences and some differences between the contours, especially due to the asymmetry of $p(u,\tau|u_{0},\tau_0)$.}
\label{singlecpdf}
\end{figure}
\end{center}
The contours of the single conditioned pdfs $p(u,\tau_0|u_{1},\tau_1)$ and $p(u,\tau_0|u_{rec},\tau_1)$ give a good impression of the quality of the reconstruction (fig.\ref{singlecpdf}) in the relation between the developments of the pdfs in time. While the overall tendency of the structure of the main peaks and their form is very similar, the spatial extension differs a bit between the two pdfs.

\subsection{Reconstructing the spatial distributions}
\begin{center}
\begin{figure}[htbp]
$\begin{array}{c@{\hspace{0.3in}}c}  
 \multicolumn{1}{l}{\mbox{\bf }}  &
 \multicolumn{1}{l}{\mbox{\bf }} \\ [-0.5cm]
\epsfxsize=2.78in
\epsffile{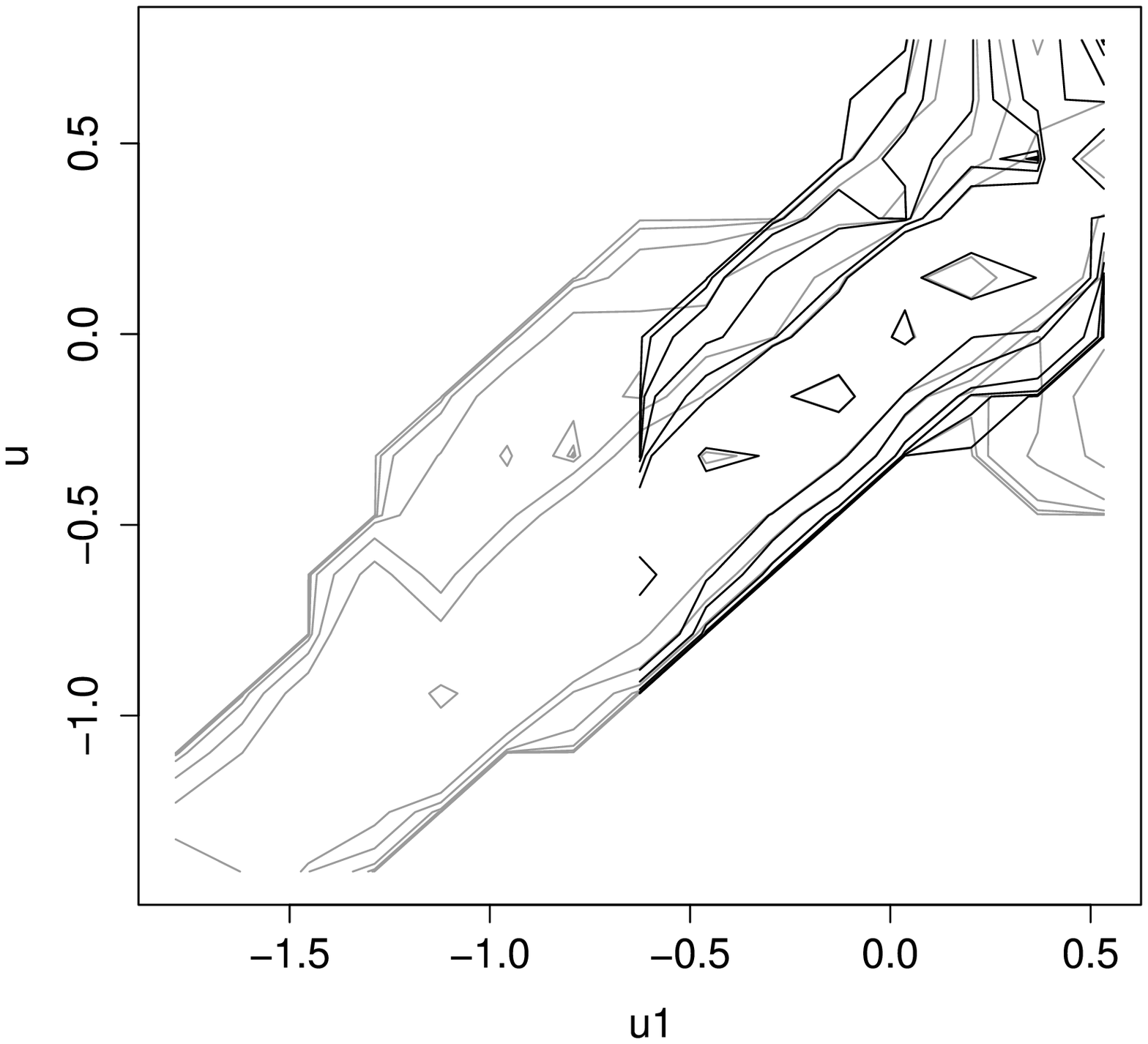}
 \put(-210,150){a)}  & 
\epsfxsize=2.78in
\epsffile{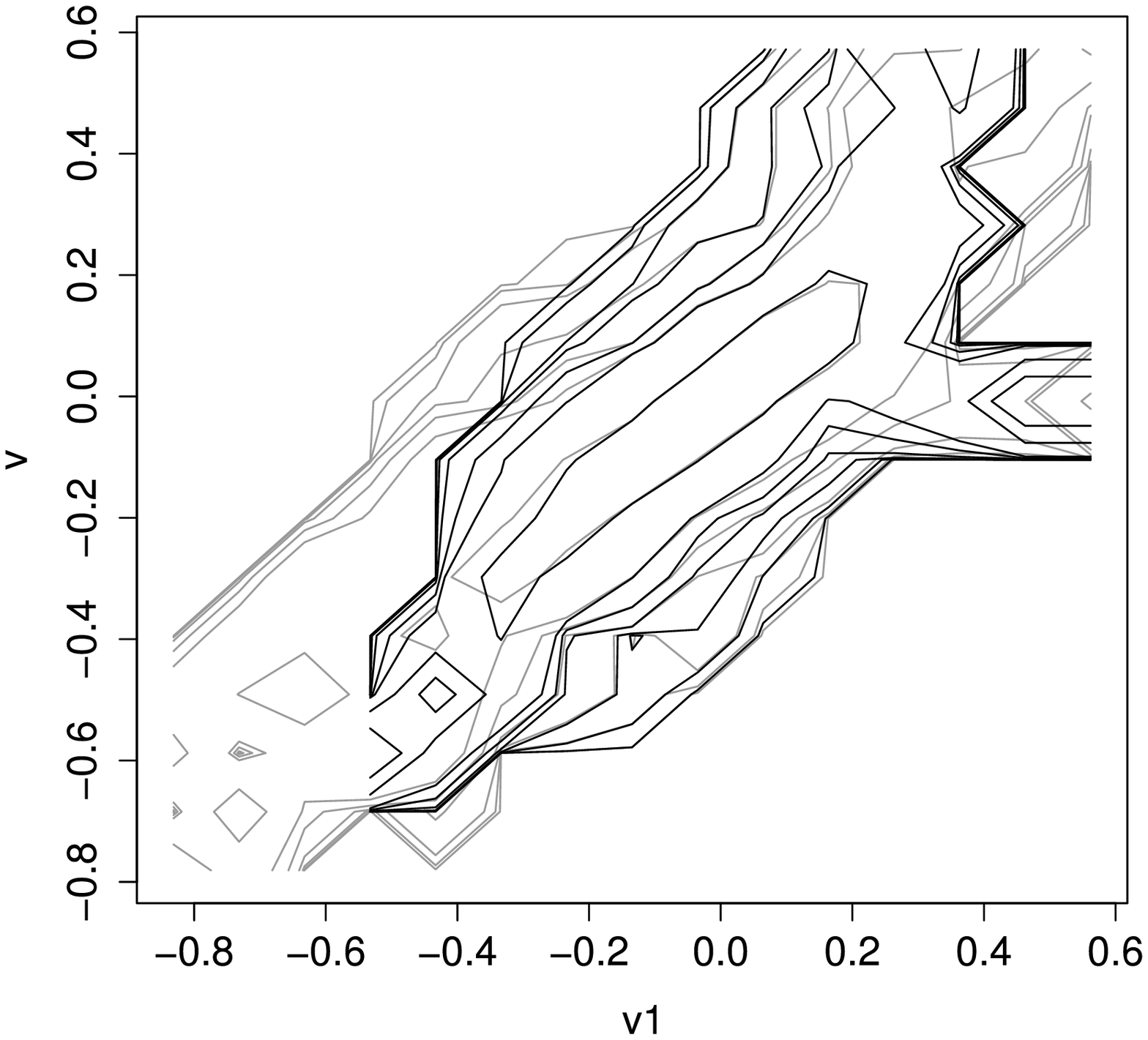}
\put(-200,150){b)}\\ [1.0cm]
\end{array}$
\caption{Contour-plot of conditioned pdfs of the time series  $p(u_i,r_0|u_{i,1},r_1)$ (grey) and  $p(u_i,r_0|u_{i,1},r_1;u_{i,2},r_2)$ (black) for the u- and v-velocity increments between point r0=d(218,61), r1=d(218,13) and r2=d(218,119).}
\label{cpdf-space}
\end{figure}
\end{center}

\begin{figure}[htb]
	\includegraphics[width=0.60\linewidth]{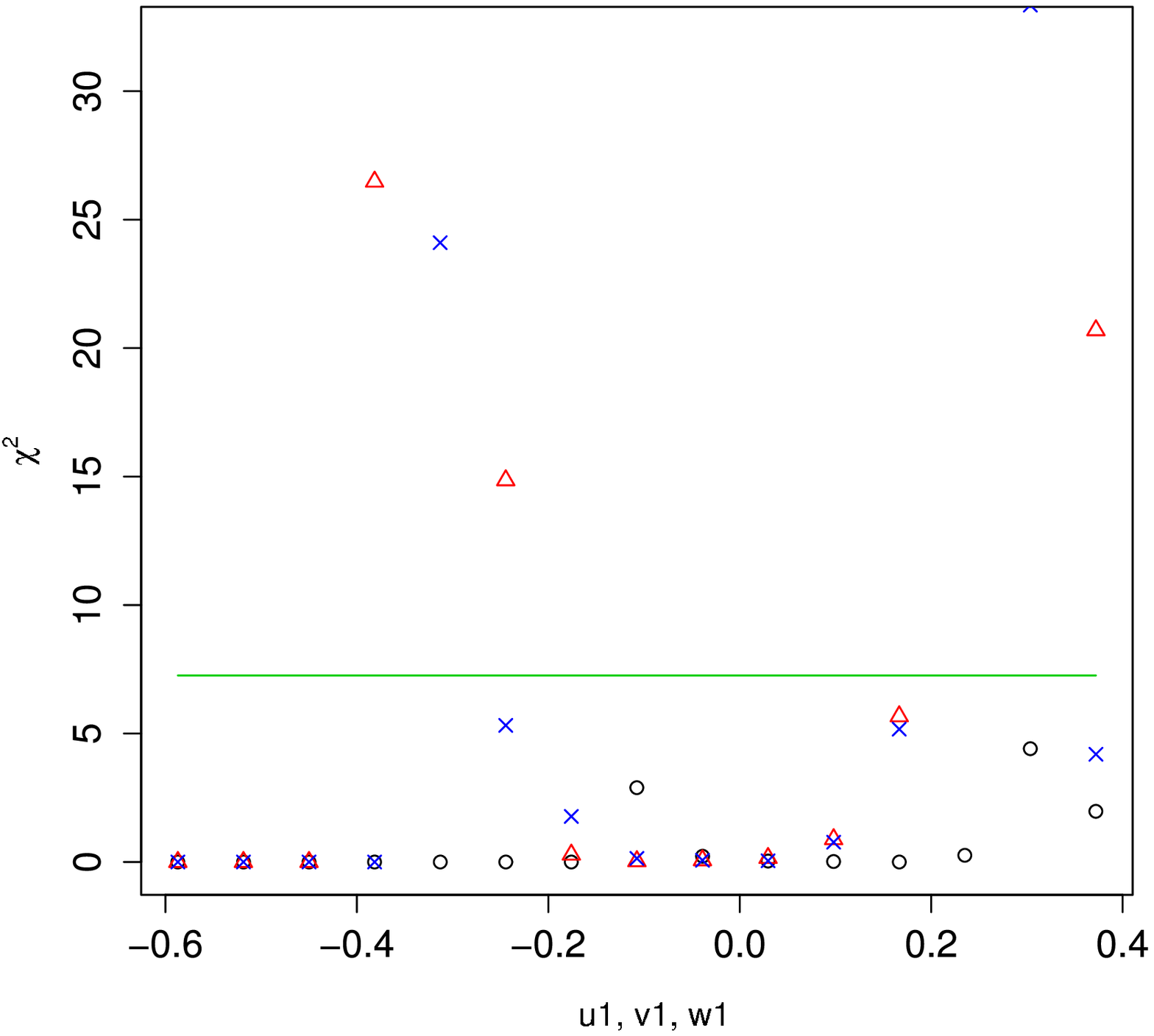}
        \caption{$\chi^2$ values for the conditioned pdfs for the increments between points 218, 61, 13 and 119 with overall scales $r=0.054$, $r_0=0.076$ and $r_1=0.088$.}
	\label{chi2-space}
\end{figure}

\begin{center}
\begin{figure}[htbp]
$\begin{array}{c@{\hspace{0.3in}}c}  
 \multicolumn{1}{l}{\mbox{\bf }}  &
 \multicolumn{1}{l}{\mbox{\bf }} \\ [-0.5cm]
\epsfxsize=2.78in
\epsffile{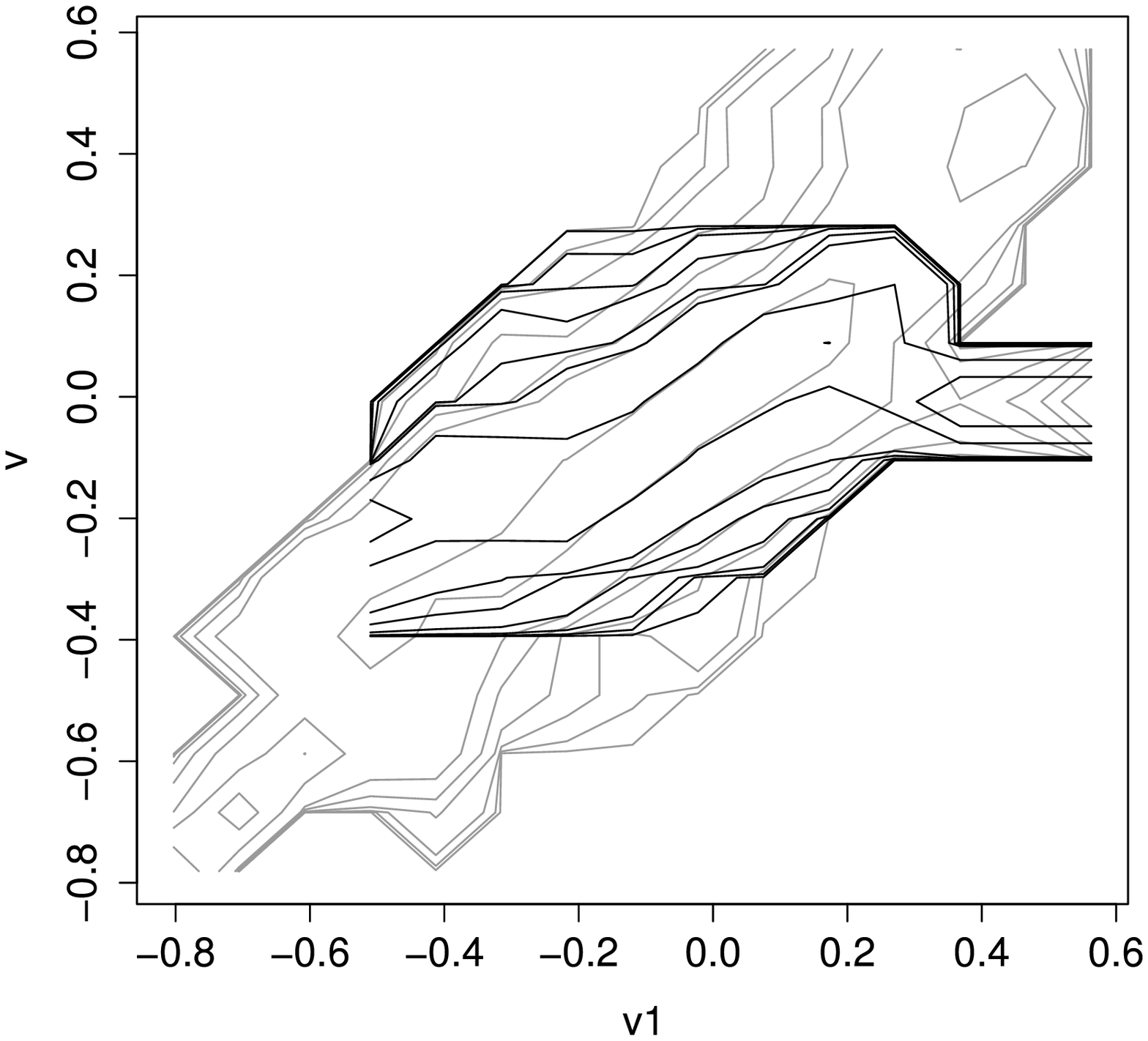}
 \put(-210,150){a)}  & 
\epsfxsize=2.78in
\epsffile{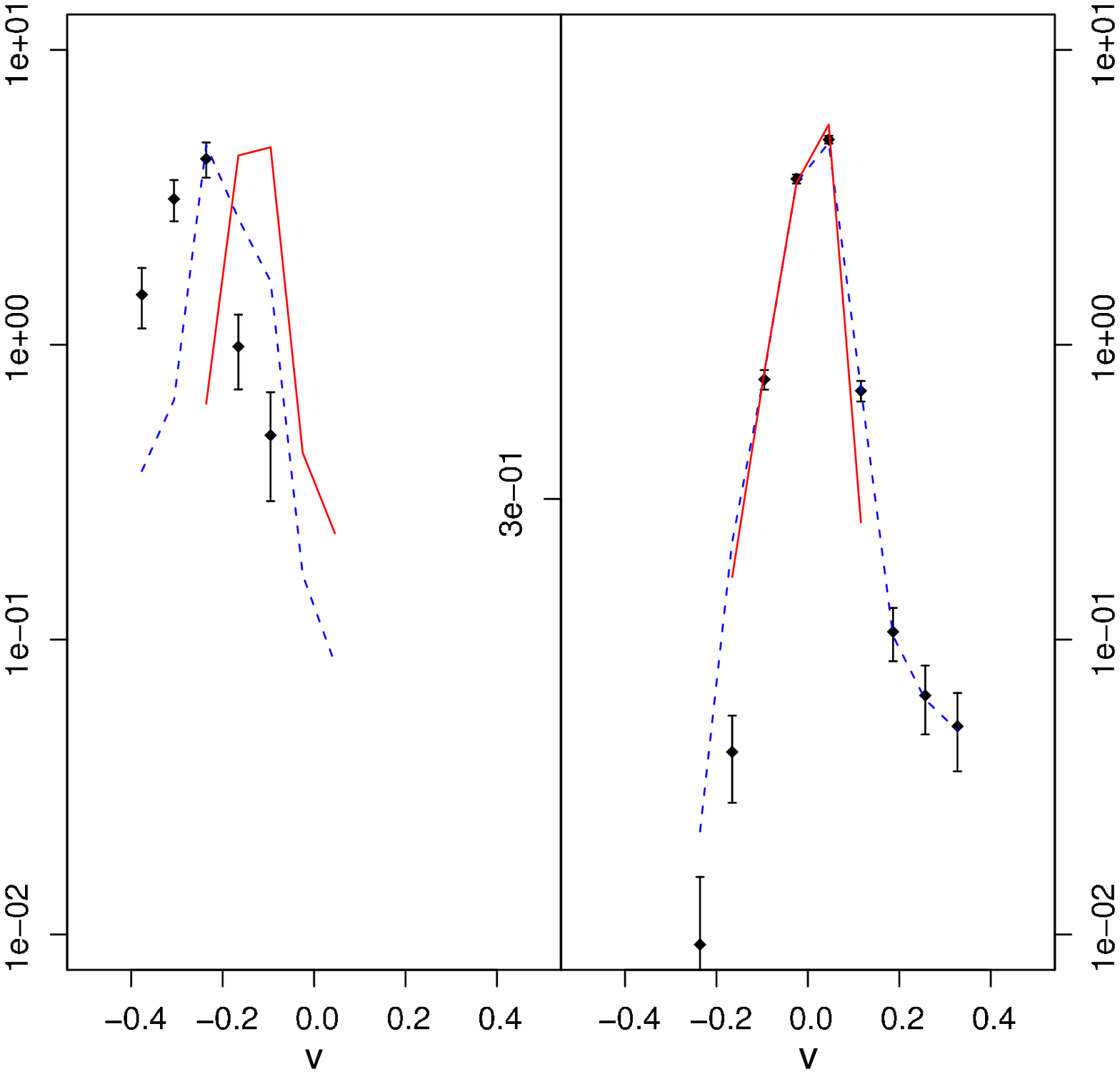}
\put(-200,150){b)}\\ [1.0cm]
\end{array}$
\caption{Contour-plot of conditioned pdfs of the time series  $p(u_i,r_0|u_{i,1},r_1)$ (grey) and  $p(u_i,r_0|u_{i,1},r_1;u_{i,rec},r_2)$ (black) for the v-increments a). Slices of the plot at  $+/-0.9\sigma$ for the v component. Here the dotted line represents the data from $p(u_i,r_0|u_{i,1},r_1;u_{i,2},r_2)$ and the solid line the data from $p(u_i,r_0|u_{i,1},r_1;u_{i,rec},r_2)$. The points with error bars are obtained from the data of the simulation with the errorbars being $\sqrt{N}$.}
\label{rec-cpdf-space}
\end{figure}
\end{center}

\begin{center}
\begin{figure}[htbp]
$\begin{array}{c@{\hspace{0.3in}}c}  
 \multicolumn{1}{l}{\mbox{\bf }}  &
 \multicolumn{1}{l}{\mbox{\bf }} \\ [-0.5cm]
\epsfxsize=2.78in
\epsffile{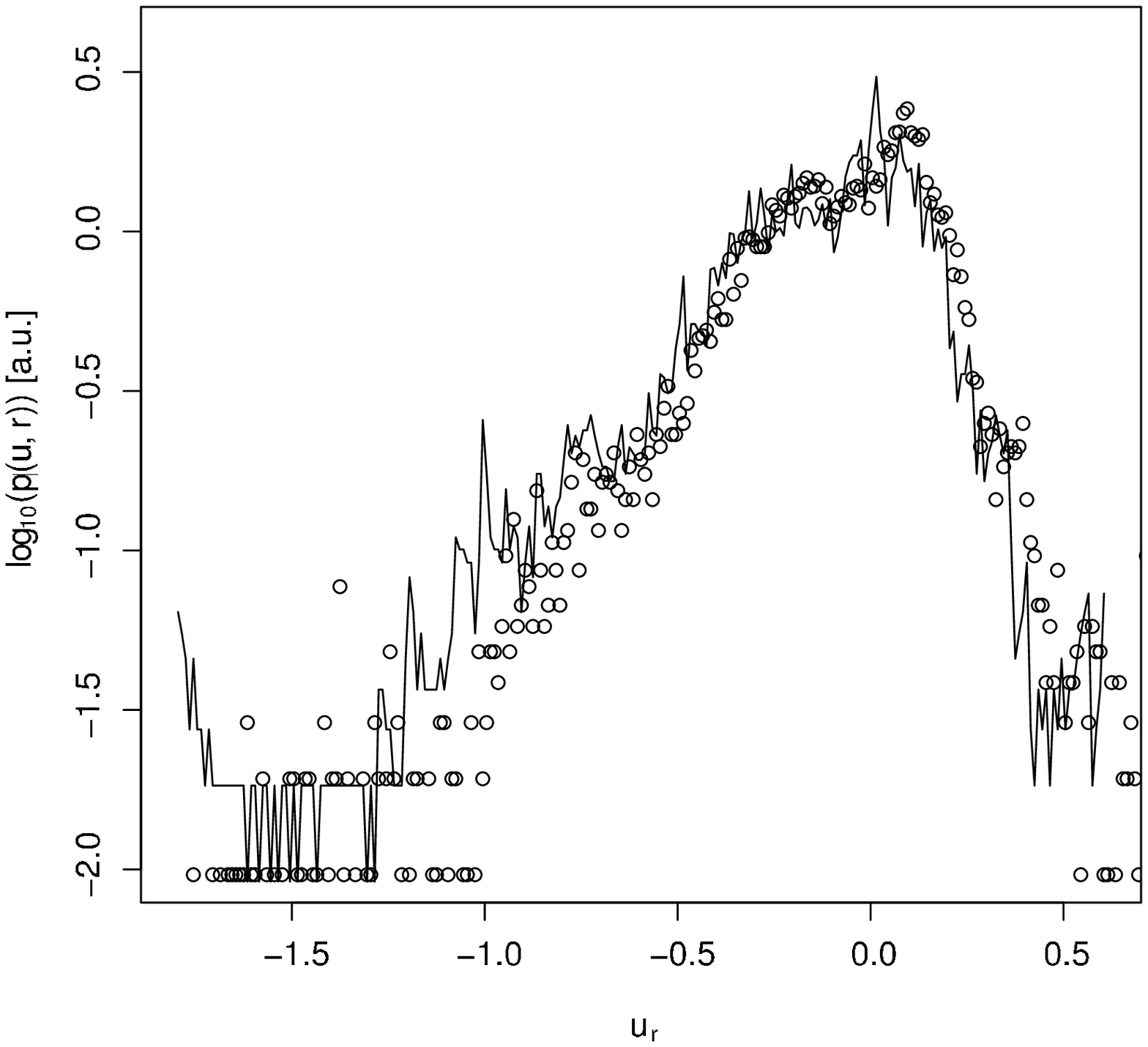}
 \put(-210,150){a)}  & 
\epsfxsize=2.78in
\epsffile{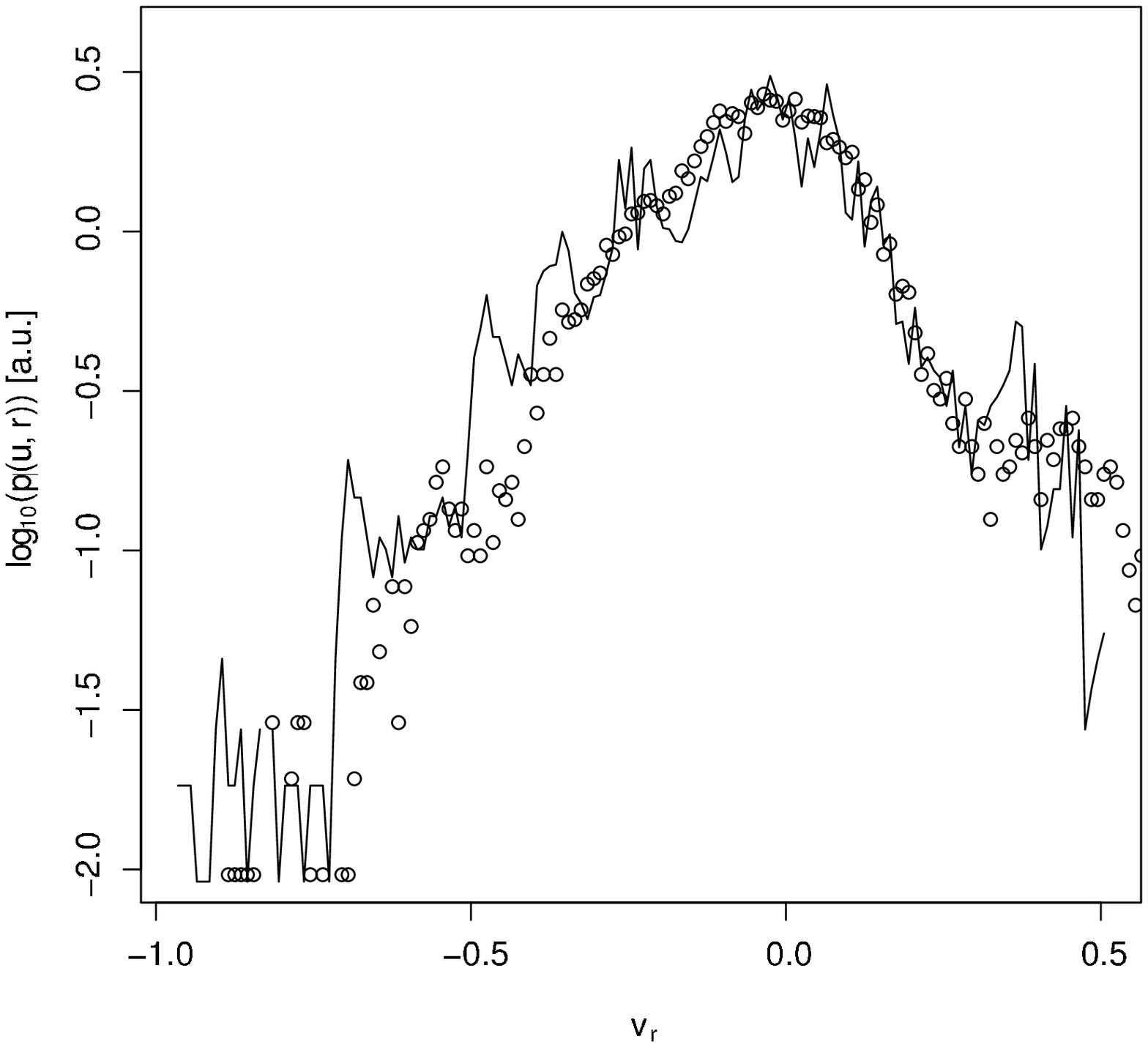}
\put(-200,150){b)}\\ [1.0cm]
\multicolumn{1}{l}{\mbox{\bf }}  &
\multicolumn{1}{l}{\mbox{\bf }}   \\ [-1cm]
\epsfxsize=2.78in
\epsffile{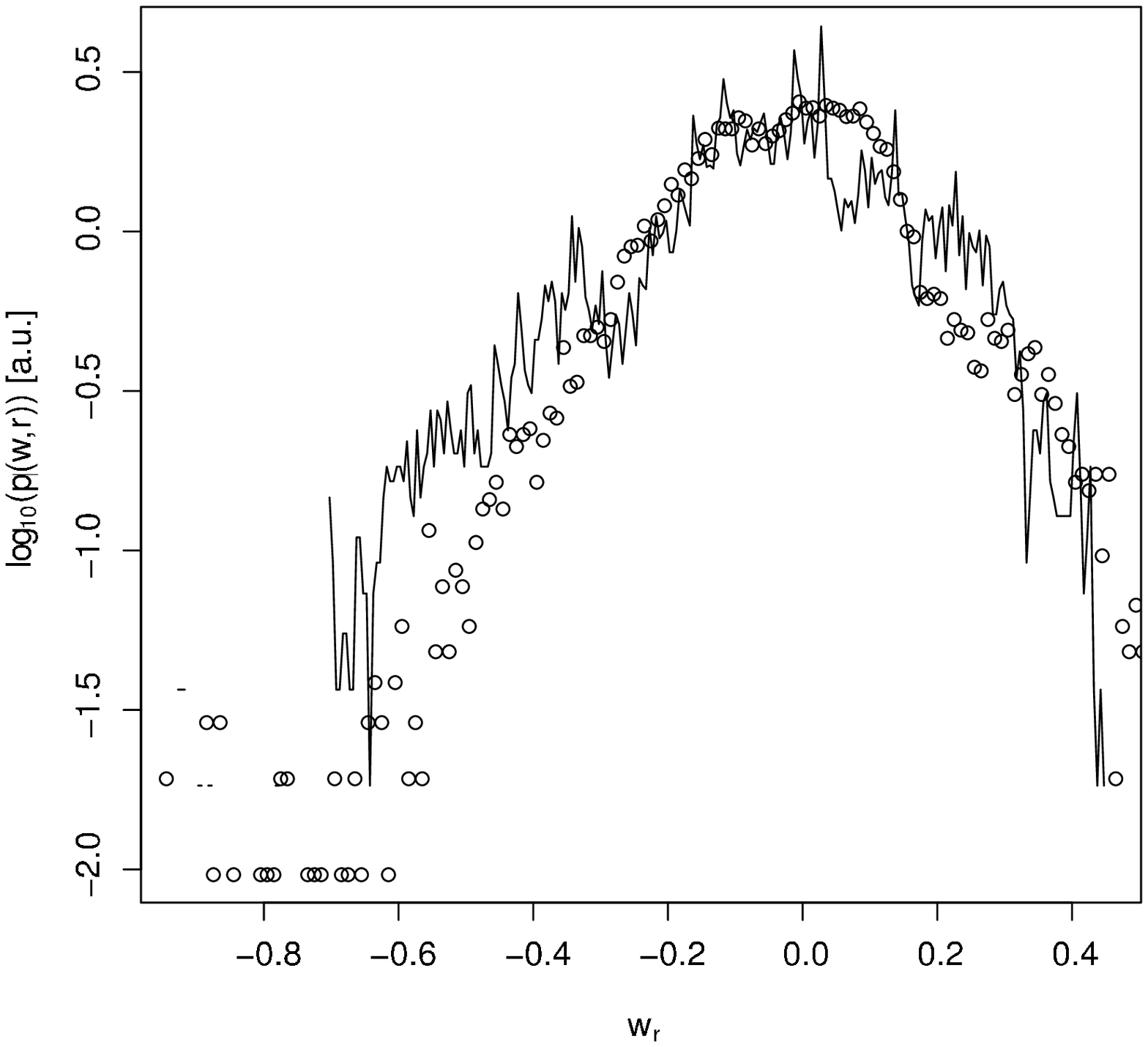}
 \put(-210,150){c)} &
\epsfxsize=2.78in
\end{array}$
\caption{The semi-logarithmic plot of the pdf of the velocity increments for all components, where the line is the distribution coming from the original simulation data and the dots represent the reconstructed time series.}
\label{pdf-spac}
\end{figure}
\end{center}

 One of the main problems for flow simulations is however the unknown flow statistics of the flow at a distant point. After having obtained convincing results for the Markov properties of velocity increments, we will apply next the Taylor hypothesis to investigate the velocity increments in for the selected points 218,61,13, and 119 of fig. \ref{points}.\\
For local isotropic turbulence it is well know that for increment statistics in space and in time are related by the Taylor hypothesis. Here we will consider a flow with a shear which is not homogeneous or local isotropic. The open question is, whether it is possible to use the above obtained characteristics of the turbulent flow for an application in space.\\
First we start with the Markov properties. Fig. \ref{cpdf-space} shows the contours of the conditioned pdfs $p(u_i,r_0|u_{i,1},r_1)$ and $p(u_i,r_0|u_{i,1},r_1;u_{i,2},r_2)$ of the u- and v-velocity increments between the points 218-61, 218-13 and 218-119, with overall distances $r_0=0.054$, $r_1=0.076$ and $r_2=0.088$.
The contours in fig.\ref{cpdf-space} indicate good Markov properties. This is reflected in the $\chi^2$ values for the two distributions shown in fig.\ref{chi2-space} where some of the outer bins have a lack of data points. However, especially close to the mean velocity increments $\overline{u}_{i,r_0}$ the Markov properties can be very well assumed as the probability of a non Markovian field is $\le 5\%$.\\
Taking the result as a strong indication that Markov properties hold we want to show that the spatial statistics can be reproduced by the stochastic processes we extracted in the previous section from the increments using a time step $\tau$.  We reconstruct from the results obtained for the point 218 data series using eq. (\ref{Langevin}). For a $\tau=\frac{r}{u}$ where $r=d(218,119)$ the correspondig conditional pdf can be obtained for all components (see fig.\ref{rec-cpdf-space}).\\
Further we compare the pdfs of the increments of $r_2=d(218,119)$ of the simulated data with the reconstructed increments. The results for all components are given in fig.\ref{pdf-spac}. As can been seen the spanwise flow direction was obviously the easiest to reconstruct. Since there was no big difference velocity magnitude between the two points and the original distribution was also symmetrical. For the pdfs of the u- and v-velocity increments show good agreement, even though slight deviations can be observed.\\
By adding the reconstructed increment to the velocity time series, one can finally resolve a reconstructed time series for the velocity at point 119:
\begin{eqnarray}
u_{i,119_{rec}}(t)=u_{i,218}(t)-u_{r2,rec}(t)
\notag
\label{reconstruct}
\end{eqnarray}
with $u_{i,218}(t)$ being the $u_i$-velocity component at the point 218 at the time $t$ and $u_{r2,rec}$ the reconstructed increment for $r_2=d(218,119)$. Thus we are able to compare the characteristics of the original velocity at point 119 given in table \ref{spacing} to a velocity reconstruction. Table \ref{rec-scales} displays the results of this reconstruction. Concerning the mean, the u-velocity shows some deviation. Although the tendency to reduce the velocity closer to the airfoil surface is correct, this reduction is not yet large enough. On the other hand the values for the v- and w-velocity are met quite well. This contrast leads to the conclusion, that mainly the u-component had effects in the flow field which were not grasp by the Gaussian statistics.
Nevertheless the results was surprisingly good as in an inhomogeneous, anisotropic field with a physical object in the area it was possible to reconstruct the field by pure statistical analysis.
\begin{table}
\begin{tabular}{|c|c|c|c|c|c|c|}
\hline
Point&Component&$\overline{u_i}$&$\sigma_i$&$L_i$&$\lambda_i$&$\eta_i$\\
\hline
119-rec&x&-0.093&0.292&0.162&0.0015&0.0001\\
\hline
119-rec&y&-0.042&0.199&0.127&0.0005&9.4e-05\\
\hline
119-rec&z&-0.002&0.192&0.115&0.0002&6.8e-05\\
\hline
\end{tabular}
\label{rec-scales}
\caption{Length scales, mean and standard deviation of the reconstructed time series at point 119.}
\end{table}

\section{Conclusions}
We analyzed a simulated flow field with the aim to gain information about the statistical properties of the field for the reconstruction of the statistics of the flow as it might be useful for stochastic models in computational fluid dynamics. Further the attempt was made to reconstruct the statistical property of the flow at one point from the stochastics at a different point. The field was obtained by a simulation around an airfoil using the spectral element code $\cal N \varepsilon \kappa \cal T \alpha$\textit{r}.\\
The simulation generated a well resolved flow field with an anisotropic and inhomogeneous turbulence area in the wake of the airfoil. The resolution of the field enabled us to use the flow field for an analysis of the stochastic characteristics of the turbulent wake using n-point correlation methods with velocity increments.\\
In a first step the time series has been analyzed at a point within an anisotropic and inhomogeneous turbulent field in the wake of the airfoil. Although the number of data points was low there was a strong indication that there exist Markov properties for time scales greater than a certain ``Markov timescale''. Applying the Taylor hypothesis this would correspond to a Markov length of $\lambda_M=0.011$ for the u-velocity component and $\lambda_M=0.018$ for the v-velocity component respectively. However due to the lack of data an uncertainty on Markov properties remains for the more rare incidences of large increments.\\
Assuming that Markov properties hold, the conditional moments $M^{(n)}(u,\tau,\Delta\tau)$ and Kramers-Moyal coefficients $D^{(n)}(u,\tau)$ for the time series for different $\Delta \tau$ and $\tau$ have been calculated. $D^{(4)} = 0$ was not given, however $D^{(4)} << D^{(2)}$ was given. Hence the further analysis has been conducted under the assumption that the conditions of Pawulas theorem were met. The first two Kramers-Moyal coefficients were used for a reconstruction of the incremental time series at the analysis position.\\
This reconstruction of the incremental time series by a Langevin equation gave promising results. Slight deviations were observed since we used Gaussian methods for reconstruction. Thus skewness and kurtosis did show some differences.\\
Further we used the obtained coefficients to estimate the structure of the flow field in real space of separated points within the inhomogeneous turbulent field by applying the Taylor hypothesis. Therefore the incremental analysis was also done using four points aligned in the wake including the original analysis point. The distance between the base point and the position of the reconstruction was about 5-8 times the Markov length scale $\lambda_M$.\\
The results were surprisingly good, especially for the v-velocity component. As the flow field is very inhomogeneous due to the shear flow on the one side of the field and a boundary layer towards the airfoil, the reconstruction just based on the statistics of the field in one point using a basic Langevin equation, the results quite good.\\
Since the results were quite promising, we do encourage further research in this area. It would be especially promising to extend the model to non Gaussian Langevin equations for the inclusion of the effect of higher order moments. It also should be investigated to which accuracy and on which scales it is possible to assume the statistics of a distant point from the statistics in another point - within an inhomogeneous filed. The results in this work give hope that such methods could lead to quite satisfying results, particularly since the dataset was still very short for the used method.\\
Towards the aim of implementing the method into a general stochastic turbulence model also further questions need to be answered, such as how the flow properties change in the boundary layer, the direct shear area or for different Reynolds numbers. Answering these questions should make the method implementable in stochastic models for flow simulations as the ones proposed by Laval or Bakosi \cite{laval2006}\cite{bakosi08} and increase their accuracy.


\subsection*{Acknowledgments}
         We would like to acknowledge Prof. S. Sherwin for letting us use the code and his support in its use.



\bibliographystyle{aiaa} 

\bibliography{Paper_Stoeve_2712}

\IfFileExists{\jobname.bbl}{}
 {\typeout{}
  \typeout{******************************************}
  \typeout{** Please run "bibtex \jobname" to optain}
  \typeout{** the bibliography and then re-run LaTeX}
  \typeout{** twice to fix the references!}
  \typeout{******************************************}
  \typeout{}
 }

\end{document}